\documentclass[10pt]{JHEP3}

\usepackage{amsmath}
\usepackage{amsfonts}
\usepackage{amssymb}
\usepackage{graphicx}
\usepackage{epsfig}
\newcommand{\comma}{\; ,}
\newcommand{\point}{\; .}
\newcommand{\SC}{\mathcal S}
\newcommand{\UC}{\mathcal U}
\newcommand{\OC}{\mathcal O}

\newcommand{\dif}{\mbox{d}}

\newcommand{\real}{{\rm Re}\hskip 1pt}

\newcommand{\Tr}{{\rm Tr}~}

\newcommand{\SU}[1]{{\rm SU(} #1 {\rm )}}

\newcommand{\LambdaR}{\Lambda_{\rm R}}
\newcommand{\LambdaUV}{\Lambda_{\rm UV}}
\title{Light scalars in strongly-coupled extra-dimensional theories.}
\author{Luigi Del~Debbio, Enrico Rinaldi\\
  SUPA and The Tait Institute, School of Physics and Astronomy, University of
  Edinburgh\\
  Edinburgh EH9 3JZ, UK\\
  E-mail: \email{luigi.del.debbio@ed.ac.uk,}
  \email{e.rinaldi@sms.ed.ac.uk}
}
\author{Alistair Hart\\
  Cray Exascale Research Initiative Europe, JCMB, King's Buildings,\\
  Edinburgh EH9 3JZ, UK\\
  E-mail: \email{a.hart@ed.ac.uk}
}
\date{\today}
\abstract { The low--energy dynamics of five--dimensional Yang--Mills
  theories compactified on $S^1$ can be described by a
  four--dimensional gauge theory coupled to a scalar field in the
  adjoint representation of the gauge group. Perturbative calculations
  suggest that the mass of this elementary scalar field is protected against
  power divergences, and is controlled by the size of the extra
  dimension $R$.  As a first step in the study of this phenomenon
  beyond perturbation theory, we investigate the phase diagram of a
  $\SU{2}$ Yang--Mills theory in five dimensions regularized on
  anisotropic lattices and we determine the ratios of the relevant physical
  scales. The lattice system shows a dimensionally
  reduced phase where the four--dimensional correlation length is much
  larger than the size of the extra dimension, but still smaller than
  the four--dimensional volume. In this region of the bare parameter
  space, at energies below $1/R$, the non--perturbative spectrum
  contains a \emph{light} scalar state. This state has a mass that is
  independent of the cut-off, and a small overlap with glueball
  operators.  Our results suggest that light scalar fields can be
  introduced in a lattice theory using compactified extra dimensions,
  rather than fine tuning the bare mass parameter.}

\preprint{}
\begin{document}
%\maketitle
%
%%%%%%% sections %%%%%%%
%
\section{Introduction}
\label{sec:intro}
Five--dimensional Yang--Mills theories compactified on a circle have a
light scalar mode, whose mass renormalization in perturbation theory
is protected by the remnant of the higher--dimensional gauge
symmetry. This light scalar is the static Kaluza--Klein mode coming
from the compactification of the fifth component of the original gauge
field, when periodic boundary conditions are
imposed~\cite{Kaluza:1921tu,Klein:1926tv}. At tree level, this mode is
massless. At low energies, the physics of compactified
extra--dimensional theories can be described by an effective
lagrangian for a four--dimensional gauge theory coupled to an
elementary scalar particle in the adjoint representation of the gauge
group. This effective description is valid only up to the
compactification energy scale $\LambdaR \sim R^{-1}$, where $R$ is the
radius of the extra dimension. At the compactification scale, other
massive vector modes become relevant for the dynamics, and their
coupling to the low--energy spectrum described above can no
longer be neglected.\\
Quantum corrections usually yield divergences in the
mass of scalar particles: in a generic renormalizable quantum field
theory, the scalar mass receives contributions proportional to the
square of the ultra--violet (UV) cut--off. However, the mass of the
scalar field coming from the compactification of a higher--dimensional
gauge field remains finite, as suggested by one--loop and two--loop
calculations
~\cite{Hosotani:1983xw,Hatanaka:1998yp,Cheng:2002iz,vonGersdorff:2002as,Hosotani:2005fk,Hosotani:2007kn}.\\
These perturbative calculations are performed using the explicit
four--dimensional effective field theory that describes the
five--dimensional system at low energies. The same result has been
obtained using the full five--dimensional gauge theory which
explicitly includes all the higher energy
contributions~\cite{DelDebbio:2008hb}. Since the extra--dimensional
gauge theory is believed to be non--renormalizable, it can only be
defined as a regulated theory with an ultra--violet cut--off
$\LambdaUV$ always in place. Interestingly the quantum
corrections to the scalar mass are independent of $\LambdaUV$:
\begin{equation}
  \label{eq:one-loop-mass}
  \delta m^2 \; = \; \frac{9 g_5^2 N_c}{32 \pi^3 R^3} \zeta (3)
  \comma
\end{equation}
where $\zeta$ is the Riemann Zeta--function, $N_c$ is the number of
colours and $g_5^2$ is the dimensionful coupling constant of the
five--dimensional Yang--Mills theory.\\
It must be stressed, that Eq.~\eqref{eq:one-loop-mass} is valid
only in the regime where there is a scale separation $\LambdaR \ll
\LambdaUV$ between the
compactification scale $\LambdaR$ and the cut--off $\LambdaUV$,
because in this case the details of the regularization can be
neglected. In this energy region, the highly energetic modes at the
cut--off scale see the extra dimension as non--compact and therefore
do not contribute to the scalar mass corrections, due to the
higher--dimensional gauge symmetry.\\ 
All the aforementioned results make the compactification mechanism a
very interesting and promising scenario to protect the mass of scalar
particles from cut--off effects. Moreover, since the early work of
Ref.~\cite{Antoniadis:1990ew}, higher--dimensional theories have
gained significant phenomenological interest.\\
In this work, we study this mechanism
in the simple case where the extra dimension is compactified on a
circle $S^1$. In particular, we would like to explore the validity of
the perturbative prediction Eq.~\eqref{eq:one-loop-mass} in the
strongly--coupled regime of the theory. While we do not expect the
proportionality constant to remain unchanged, we want to check whether
the non--perturbative dynamics preserves the independence of the UV
cut--off, and the functional dependence on the compactification
radius. This is a non-trivial task, since in the non--perturbative
regime the states in the spectrum are not the excitations of the
elementary fields in the action.\\
To be able to access the non--perturbative regime, we use Monte Carlo
simulations of a lattice gauge theory in five dimensions. We then
search for a region in the parameter space of the lattice theory where
the hierarchy of scales is such that the low--energy physics is
described by a four--dimensional effective theory with a light
scalar particle.\\
In recent years, there have been several numerical studies of the simplest
of these extra--dimensional theories on the lattice, namely the $\SU{2}$
pure gauge theory on a five--dimensional torus with anisotropic
lattice spacings, $a_4$ in the four--dimensional space and $a_5$ in
the extra fifth
dimension~\cite{Ejiri:2000fc,Ejiri:2002ww,deForcrand:2010be,Knechtli:2010sg,Farakos:2010ie}.
A pioneering study of the same model on isotropic lattices was done in
the late seventies~\cite{Creutz:1979dw}.\\ 
The aim of this work is to explore the parameter space of the lattice
model and to define the scales separation by studying the behaviour of
observables such as the string tension and the mass of scalar states;
this goes in the direction of improving previous recent 
results~\cite{deForcrand:2010be} and trying to clarify the status of
Eq.~\eqref{eq:one-loop-mass} in the non--perturbative regime. Using
numerical simulations in the region of phase space where there is a
hierarchy of scales $\LambdaR \ll \LambdaUV$, we are able to study the
parametric dependence of the non--perturbative scalar mass on the
cut--off $\LambdaUV$ and on the compactification scale
$\LambdaR$. However, in order to fully understand the nature of the
effective theory and of the scalar particle, more studies are needed
that are beyond the scope of this work. In particular, matching
simulations between this five--dimensional gauge model and the
four--dimensional gauge theory with an adjoint scalar field in the
action could be performed, following what was done to test dimensional
reduction in lower dimensions~\cite{Hart:1999dj}.\\ 
In Sec.~\ref{sec:lattice-setup} we describe the lattice setup used in
our simulations of the $\SU{2}$ Yang--Mills theory in five
dimensions. In Sec.~\ref{sec:scale-separation} we explain the
separation of scales that we expect to find in the lattice model, and
we analyse the perturbative predictions for the behaviour of the
desired hierarchy of scales as we scan the bare parameter space.
Next, we provide a description of the phase diagram of the model in
Sec.~\ref{sec:phase-diagram} and compare our findings with previous
studies. Once the phase diagram has been mapped out and the
interesting region has been identified, we present the details of our
measurements and compare them with the perturbative expectations in
Sec.~\ref{sec:results}. Finally we present a critical discussion of
the results and future developments of these ideas.

%
%%%%%%%%%%%%%%%%%%%%%%%%%%%%%%%
%
\section{The lattice model}
\label{sec:lattice-setup}
The continuum, five--dimensional pure gauge theory is defined by the
Euclidean action
\begin{equation}
  \label{eq:continuum-action}
  \SC \; = \; \int \dif^4 x \: \int_0^{2\pi R} \dif x_5 \:
  \frac{1}{2g_5^2} \Tr F_{MN}^2
  \comma
\end{equation}
where periodic boundary conditions are imposed along the fifth
direction (whose coordinate is $x_5$) in order to make it compact.
The field--strength tensor is the extra--dimensional
generalization of the four--dimensional one
\begin{equation}
  \label{eq:field-tensor}
  F_{MN} \; = \; \partial_M A_N - \partial_N A_M + i [A_M,A_N]
\quad M,N=1,\dotsc,5 \point
\end{equation}
This continuum theory has an infinite four--dimensional volume, but it
is defined only on a finite and compact fifth dimension of
length $L_5 = 2 \pi R$, where $R$ is the compactification radius.\\
Since this theory is perturbatively non--renormalizable, the
ultra--violet cut--off $\LambdaUV$ cannot be removed. For the same
reason we consider the action in Eq.~\eqref{eq:continuum-action} only
as the simplest non--trivial example of effective theory in five
dimensions: an arbitrary number of operators and couplings could be
added in principle. Cut--off effects
are expected to be irrelevant in the low--energy regime of the
theory defined by the action
in Eq.~\eqref{eq:continuum-action}, i.e. at scales $E \ll \LambdaUV$.\\
The continuum action is regularized on a five--dimensional lattice,
where the finite lattice spacing determines the shortest propagating
wavelength. Two independent lattice spacings $a_4$ and $a_5$ can be defined on the
lattice, which correspond respectively to the lattice spacing in the
four--dimensional subspace, and in the extra fifth direction; the
bigger of the two defines the inverse of the cut--off $\LambdaUV$. The
gauge potential $A_M(x_M)$ is replaced on the lattice by gauge link
variables $\UC_M(x)$ joining the site $x$ and the site $x + a\hat{M}$,
where $a = a_4$ if $M=1,2,3,4$ and $a = a_5$ if $M=5$. Periodic
boundary conditions for the gauge
links are imposed in all five directions.\\
We choose the \emph{anisotropic} lattice Wilson action for $\SU{N_c}$
gauge theories:
\begin{equation}
  \label{eq:aniso-lattice-action-E}
  \SC_W \; = \; \beta_4 \sum_{x;1\leq \mu <
    \nu \leq 4} \left[ 1 - \frac{1}{N_c} \real \Tr P_{\mu\nu}(x)
  \right] +
  \beta_5 \sum_{x;1\leq \mu \leq 4}
  \left[ 1 - \frac{1}{N_c} \real \Tr P_{\mu 5}(x)
  \right] 
  \comma
\end{equation}
where $P_{\mu \nu}$ is the four--dimensional plaquette ($\mu$
and $\nu$ run from $1$ to $4$)
\begin{equation}
  \label{eq:4d-plaq}
    P_{\mu\nu}(x) \; = \; \UC_{\mu}(x)\UC_{\nu}(x+\hat{\mu}a_4)\UC_{\mu}^\dag(x+\hat{\nu}a_4)
  \UC_{\nu}^\dag(x)
  \comma
\end{equation}
and $P_{\mu 5}$ is the
plaquette abutting on an extra--dimensional slice
\begin{equation}
  \label{eq:5-plaq}
    P_{\mu5}(x) \; = \; \UC_{\mu}(x)\UC_{5}(x+\hat{\mu}a_4)\UC_{\mu}^\dag(x+\hat{5}a_5)
  \UC_{5}^\dag(x)
  \point
\end{equation}
The sum is intended to be on all the lattice sites $x$ of the full five
dimensional lattice volume.\\
This lattice setup is the same used
in Ref.~\cite{Ejiri:2000fc}. However, a
different parametrization for the Wilson action can be
used~\cite{deForcrand:2010be}:
\begin{equation}
  \label{eq:aniso-lattice-action-P}
  \SC_W \; = \; \frac{\beta}{\gamma} \sum_{x;1\leq \mu \leq
    \nu \leq 4} \left[ 1 - \frac{1}{N_c} \real \Tr P_{\mu\nu}(x)
  \right] +
  \beta \gamma \sum_{x;1\leq \mu \leq 4}
  \left[ 1 - \frac{1}{N_c} \real \Tr P_{\mu 5}(x)
  \right] 
  \comma
\end{equation}
where the lattice coupling constant is
\begin{equation}
  \label{eq:beta}
  \beta \; = \; \sqrt{\beta_4 \beta_5}
  \comma
\end{equation}
and the second parameter is the bare anisotropy
\begin{equation}
  \label{eq:gamma}
  \gamma \; = \; \sqrt{\frac{\beta_5}{\beta_4}}
  \point
\end{equation}
The bare anisotropy is related to the ratio of the lattice spacings
$\xi = a_4/a_5$. At tree level $\gamma=\xi$, but quantum corrections
make $\xi$ deviate from this value. The relation between $\xi$ and
$\gamma$ for this action has already been studied in bare parameter
space and a useful map relating this two
quantities can be found in Ref.~\cite{Ejiri:2000fc}.\\
In order to obtain Eq.~\eqref{eq:continuum-action} in the classical
continuum limit of the action in Eq.~\eqref{eq:aniso-lattice-action-E},
we must require the following relations for the
lattice parameters (coupling constants) $\beta_4$ and $\beta_5$:
\begin{eqnarray}
  \label{eq:beta4-beta5}
    \beta_4 & = & \frac{2N_ca_5}{g_5^2} \\
    \beta_5 & = & \frac{2N_ca_4^2}{g_5^2a_5} \point
\end{eqnarray}
Similarly, for the action in Eq.~\eqref{eq:aniso-lattice-action-P} we
have
\begin{equation}
  \label{eq:lattice-coupling}
   \beta \; =\; \frac{2N_c}{g_5^2} a_4
   \comma
\end{equation}
 and
\begin{equation}
  \label{eq:gamma-xi}
  \gamma \; = \; \xi \; = \; \frac{a_4}{a_5}
  \point
\end{equation}
In this work we use the Wilson action
Eq.~\eqref{eq:aniso-lattice-action-E} with $\beta_4$ and $\beta_5$ as
bare parameters, and therefore our results will be presented as
functions of these two quantities. However some of the features of the
phase diagram are better explained in terms of $\beta$ and $\gamma$,
in particular when comparing our findings
to existing results~\cite{deForcrand:2010be}.\\
Finally there are two more parameters in the lattice model that can be
adjusted in order to realize the desired separation of scales; they
are $N_4$, the number of lattice sites in each of the usual four
directions, and $N_5$, the number of lattice sites in the extra
dimension. Together with the corresponding lattice spacings, they
determine the \emph{physical} size of the system: $L_4 = a_4 N_4$ in
four dimensions and $L_5 = 2 \pi R = a_5 N_5$ in the fifth dimension.\\
In the following we restrict ourself to the non--Abelian gauge
group $\SU{2}$, thus setting $N_c=2$ in the above
definitions.

%
%%%%%%%%%%%%%%%%%%%%%%%%%%%%%%%
%
\section{Dimensional reduction and scale separations}
\label{sec:scale-separation}
We have already mentioned that the theory described by the action in
Eq.~\eqref{eq:continuum-action} is perturbatively non--renormalizable
because the five--dimensional coupling constant $g_5^2$ has negative
mass dimension $M^{-1}$. Moreover, the theory possesses another
intrinsic scale when the extra dimension is compactified on the
circle: the compactification scale $\LambdaR \sim R^{-1}$. This scale
is the analogue of the temperature scale in the formulation of
finite--temperature field theories compactified on a circle.\\
Upon compactification, the gauge fields are decomposed into Fourier
modes (called Kaluza--Klein modes in this context, or Matsubara modes
in finite--temperature field theories). At the classical level the
spectrum of the theory contains massless vectors, coming from the
gauge field components in the four--dimensional subspace, and a
massless scalar, coming from the gauge component in the extra compact
direction. All the higher modes acquire masses proportional to
$\LambdaR$. In the quest for an effective description of the
low--energy physics of the theory, one can integrate out the states at
energies greater than the compactification scale, leaving a
four--dimensional gauge field coupled to an adjoint massless
scalar. However, this dimensional reduction is a sensible
description only if there is a scale separation $\LambdaR \ll
\LambdaUV$: the physics of the compactified theory is not affected by
the details of the regularization. As discussed below, this condition
is only satisfied in a specific region of the lattice bare parameter
space.\\
If we focus on the low--energy $E \ll \LambdaR$ and weakly--coupled
regime, we expect a perturbative spectrum, where the elementary scalar
particle acquires a mass through radiative corrections, while the
gauge vectors remain massless. As we explore more strongly--coupled
regimes, the theory develops a non--perturbative mass gap related to
confinement. Our aim is to study what happens to the low--lying
spectrum of scalar particles in this non--perturbative regime.  In
particular we would like to understand if there exists a region in the
bare parameter space of the five--dimensional theory where the
non--perturbative dynamics can be described by a four--dimensional
effective gauge theory coupled to a light adjoint scalar, whose mass
is decoupled from the cut--off scale as suggested by the one--loop
equation Eq.~\eqref{eq:one-loop-mass}. Moreover, it would be
interesting to find a region where the scalar mass is of the order of
the mass gap in the gauge sector.\\ 
A previous study has shown that there is a region of the phase diagram
of the lattice model where a scale separation between the static modes
of the four--dimensional gauge fields and their higher Kaluza--Klein
modes is observed~\cite{deForcrand:2010be,Kurkela:2009dg}, indicating
that the theory undergoes dimensional reduction similar to the case of
four--dimensional hot gauge theories~\cite{% Karsch:1994af,
  Datta:1998eb}. However, in that same region, the static mode of the
fifth component of the gauge field appears to be completely decoupled,
with a mass at the scale of the cut--off, and hence outside the regime
of validity of Eq.~\eqref{eq:one-loop-mass}.\\
Let us summarize now the hierarchy of scales that we would like to
find non--perturbatively in the lattice theory described in
Sec.~\ref{sec:lattice-setup}.
\begin{itemize}
\item A separation between the compactification scale and the cut--off
  \begin{equation}
    \label{eq:separation-cutoff-radius}
    \frac{\LambdaUV}{\LambdaR} \; \gg \;  1
    \point
  \end{equation}
\item The mass gap identified by the string tension $\sqrt{\sigma}$ in
  four dimensions must be separated both from the cut--off and from
  the compactification scales:
  \begin{equation}
    \label{eq:separation-sigma-cutoff}
    \sqrt{\sigma} \; \ll \; \LambdaUV \; {\rm ;} \qquad  \sqrt{\sigma} \; \ll \; \LambdaR
    \point
  \end{equation}
  In fact, only if the above relations are true, we expect the long
  distance physics to be independent of the actual regularization of
  the theory, and not to be sensitive to contributions from higher
  modes. In other words, since the string tension gives the inverse of
  the four--dimensional correlation length, when $\sqrt{\sigma}$ is
  small compared to the cut--off, then the characteristic length of
  the system is much larger than the lattice spacing, and the details
  of the discretization of the theory should become insignificant.
\item Light scalar states should be at the energy scale defined by the
  mass gap, and hence their mass, generically
  referred to as $m_5$, should be separated from the cut--off and from
  the energy scale of the other Kaluza--Klein massive modes:
  \begin{equation}
    \label{eq:separation-mass-cutoff}
    m_5 \; \simeq \; \sqrt{\sigma}\; {\rm ;} \qquad m_5 \; \ll \;
    \LambdaUV \; {\rm ;} \qquad m_5 \; \ll \; \LambdaR \point
  \end{equation}
\item Finally, we need to check the dependence of the scalar mass from
  the cut-off and the compactification radius. We would like to find a
  region in the space of bare parameters, where we have a scaling
  similar to the one Eq.~\eqref{eq:one-loop-mass} obtained for an
  elementary scalar particle in the weakly-coupled regime:
  \begin{equation}
    \label{eq:light-scalar}
    m_5^2 \; \propto \; \frac{g_5^2}{R^3}
    \point
  \end{equation}
\end{itemize}
In a strongly--coupled theory the different energy scales described
above are dynamically generated, and need to be measured by numerical
simulations. In the following discussions, we choose to express every
scale in units of the four--dimensional string tension
$\sqrt{\sigma}$; hence the other three scales in the theory are
characterized by three dimensionless ratios. The ultra--violet
cut--off $\LambdaUV$, given by the inverse of the largest lattice
spacing of the model, is
\begin{equation}
  \label{eq:cutoff-lattice}
  \frac{\LambdaUV}{\sqrt{\sigma}} \; \equiv \; \frac{1}{a_4\sqrt{\sigma}}
  \comma
\end{equation}
because we will be dealing with anisotropies $\xi = \frac{a_4}{a_5}
\geq 1$. Similarly, the compactification scale $\LambdaR$ is
\begin{equation}
  \label{eq:radius-lattice}
  \frac{\LambdaR}{\sqrt{\sigma}} \; \equiv \; \frac{1}{2 \pi R\sqrt{\sigma}} \; = \; \frac{1}{a_5 N_5 \sqrt{\sigma}}
  \point
\end{equation}
Finally, the scalar mass $m_5$ can be expressed as the ratio of the
scalar mass and the string tension both measured in units of the
lattice spacing $a_4$ in our simulations:
\begin{equation}
  \label{eq:scalar-lattice}
  \frac{m_5}{\sqrt{\sigma}} \; = \; \frac{a_4m_5}{a_4\sqrt{\sigma}}
  \point
\end{equation}
In Fig.~\ref{fig:scales} we summarize pictorially the scale separations in the theory.\\
\FIGURE[ht]{
  \begin{picture}(300,40)
    \thinlines
    \multiput(80,20)(5,0){7}{\line(1,1){10}}
    \thicklines
    \put(50,25){\vector(1,0){200}}
    \put(100,18){\line(0,1){15}}
    \put(95,0){$1$}
    \put(110,40){$\displaystyle \frac{m_5}{\sqrt{\sigma}}$}
    \put(150,20){\line(0,1){10}}
    \put(145,5){$\displaystyle \frac{1}{2\pi R \sqrt{\sigma}}$}
    \put(220,20){\line(0,1){10}}
    \put(215,5){$\displaystyle \frac{1}{a_4 \sqrt{\sigma}}$}
    \put(260,20){$\displaystyle \frac{E}{\sqrt{\sigma}}$}
  \end{picture}
  \label{fig:scales}
  \caption{The figure shows the desired separation of energy scales.
    The scales are all expressed in terms of the four--dimensional
    string tension that characterizes the low--energy physics of the
    theory. The region of energies where we expect the scalar mass to
    lie is highlighted by the shaded band.}  }
Let us note that the separation of the UV and the compactification
scales can be entirely expressed in terms of bare parameters of the
lattice model at tree level:
\begin{equation}
  \label{eq:separation-lattice}
  \frac{\LambdaUV}{\LambdaR} \; \equiv \; \frac{a_5 N_5}{a_4} \; = \; \frac{N_5}{\xi} \; \sim \; \frac{N_5}{\gamma}
  \comma
\end{equation}
where the last step follows from Eq.~\eqref{eq:gamma-xi} and is a
valid approximation only in the weak--coupling limit. However, the
scalar mass in Eq.~\eqref{eq:scalar-lattice} must be measured
non--perturbatively, because it would be divergent in the perturbative
regime ($\sqrt{\sigma}=0$).\\ 
The three energy scales of the system, $\LambdaUV$, $\LambdaR$ and
$m_5$ can be studied by adjusting the three bare parameters of the
lattice model $\beta_4$, $\beta_5$ and $N_5$ (here $N_4$ must be large
enough that the four--dimensional subspace can be considered in the
infinite volume limit). Fixing a point in the space
($\beta_4$,$\beta_5$,$N_5$), or equivalently ($\beta$,$\gamma$,$N_5$),
will dynamically determine the two lattice spacings $a_4$ and $a_5$,
together with the extent of the extra dimension $a_5N_5$. Therefore,
measuring the three scales with lattice simulations at different points
of this bare parameter space is a powerful tool to explore the
dependence of the scalar mass on $a_4$ and $R$. In particular, the
ability to investigate separately these functional dependences is the
major breakthrough of this work: contrary to what was done in
Ref.~\cite{deForcrand:2010be}, where the separation $\frac{2\pi
  R}{a_4}$ was kept fixed in the numerical simulations, we explore a
region of the phase space where $R$ and $a_4$ vary independently and
we are able to follow, non--perturbatively, lines of constant scalar
mass.\\ 
In order to gain some insight in the behaviour of the lines of
constant physics for this model, we can use perturbative results as a
guide, with the caveat that they are expected to provide a sensible
description of the data only in the weak--coupling regime. From the
one--loop renormalization group equation of a four--dimensional
Yang--Mills theory, we expect the asymptotic scaling relation
\begin{equation}
  \label{eq:string-4d}
  \sigma \; \sim \; \frac{1}{(2 \pi R)^2}
  \exp\left\{-\frac{1}{b_0 g_4^2(\LambdaR)}\right\}
  \comma
\end{equation}
where $b_0$ is the first term in the perturbative $\beta$--function of
the four--dimensional theory ($b_0=11/24\pi^2$ for $\SU{2}$) and
$g_4^2(\LambdaR)$ is the effective dimensionless coupling constant at
the compactification scale. In terms of the lattice parameters of the
model, we can rewrite Eq.~\eqref{eq:string-4d} as 
\begin{equation}
  \label{eq:string-4d-latt}
  a_4^2 \sigma \; \sim \; \frac{\gamma^2}{N_5^2}
  \exp\left\{-\frac{\beta N_5}{2 N_c b_0 \gamma}\right\}
 \point
\end{equation}
Notice that Eq.~\eqref{eq:string-4d-latt} is obtained by trading
$g_4^2(\LambdaR)$ for the lattice tree--level relation in
Eq.~\eqref{eq:lattice-coupling}, evaluated at the compactification
energy scale $\LambdaR$ (at tree level $g_4^2(\LambdaR)=
\frac{g_5^2(\LambdaR)}{2 \pi R}$). This asymptotic behaviour has been
checked numerically on the lattice in a particular region of the
parameter space of the model and in the limit $a_5 \rightarrow
0$~\cite{deForcrand:2010be}. Furthermore, if we assume the scalar mass
to behave perturbatively according to Eq.~\eqref{eq:one-loop-mass} in
the dimensionally reduced theory, we have the following expression for
the mass $m_5$ in units of the lattice spacing $a_4$
\begin{equation}
  \label{eq:m2a2}
  (m_5 a_4)^2 \; = \; \frac{2 N_c \gamma^3}{\beta N_5^3}
  \point
\end{equation}
The latter equation can be divided by Eq.~\eqref{eq:string-4d-latt} to
express the mass in units of the string tension:
\begin{equation}
  \label{eq:m2-sigma}
  \frac{m_5}{\sqrt{\sigma}} \; \sim \; \sqrt{\frac{2 N_c \gamma}{\beta
      N_5}} \exp\left\{\frac{\beta N_5}{4 N_c b_0 \gamma}\right\}
  \point
\end{equation}
 We can therefore plot the perturbative predictions from
 Eq.~\eqref{eq:string-4d-latt} and Eq.~\eqref{eq:m2-sigma} in the plane
 ($\beta, N_5/\gamma$). This is shown in Fig.~\ref{fig:perturbative},
 where some isosurfaces are labelled in order to understand the
 functional behaviour. When these perturbative formulae are used, 
 the scalar mass in units of the string tension
 has a minimum value in the bare parameter space. Moreover, in this
 weak--coupling limit, the scalar mass is always above the scale set by
 the string tension therefore decoupling from the low--energy theory
 ~\cite{deForcrand:2010be}.
\FIGURE[ht]{
  \epsfig{figure=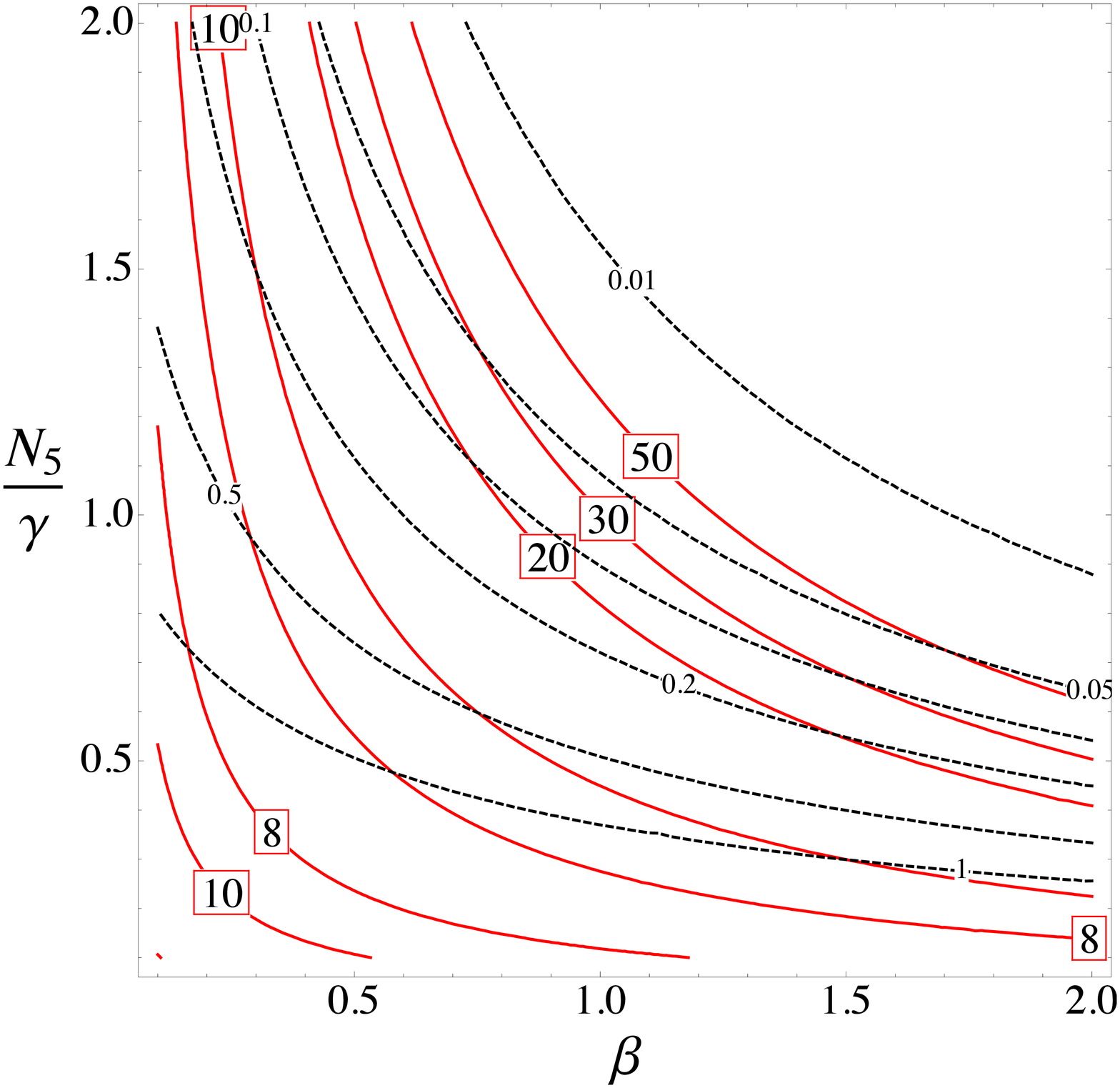,width=0.65\textwidth,clip}
  \label{fig:perturbative}
  \caption{The lines of constant lattice spacing in units of the
    string tension $a_4\sqrt{\sigma}$ are shown as dashed black lines
    in the plane ($\beta,N_5/\gamma$). The lines of constant scalar
    mass in units of the string tension $m_5/\sqrt{\sigma}$ are shown
    with solid red lines. This plot is similar to the drawing in Fig.7
    of Ref.~\cite{deForcrand:2010be}, where the authors also took into
    account the features of the phase diagram which we will describe
    in Sec.~\ref{sec:phase-diagram} (note that the coordinate
    $\beta$ is called $\beta_5$ in Ref.~\cite{deForcrand:2010be}, and
    $N_5/\gamma$ is called $\tilde{N}_5$).}
}
Keeping this in mind and assuming that Fig.~\ref{fig:perturbative}
represents the actual lines of constant physics, we
can speculate about how to reach a continuum limit for this
lattice model.  As it was firstly noted in Ref.~\cite{deForcrand:2010be}, two
different four--dimensional continuum theories can be described as the
lattice spacing $a_4$ vanishes. The one we are interested in for this
study is a $\SU{2}$ Yang--Mills theory coupled to an adjoint scalar
field: this theory is described by the lattice model following a line
of constant $m_5/\sqrt{\sigma}$ (one of the solid red lines in
Fig.~\ref{fig:perturbative}) towards smaller values of $\beta$ and
bigger values of $N_5/\gamma$. In this direction, $a_4\sqrt{\sigma}$
decreases, while the scalar mass is kept fixed, and the effects of the
regularization are suppressed by powers of $a_4\sqrt{\sigma}$. A
remarkable feature of this approach to the continuum limit is that the
direction to be taken in the bare parameter space goes towards higher
values of $N_5/\gamma$. This means that the size of the extra
dimension $2\pi R$ increases in units of the lattice spacing $a_4$,
while the theory dimensionally reduces to four dimensions as already
suggested in the D--theory non--perturbative approach to quantum field
theories~\cite{Chandrasekharan:1996ih,Brower:1997ha}.\\
Let us stress again that Eq.~\eqref{eq:string-4d-latt} to
Eq.~\eqref{eq:m2-sigma} are found using one--loop continuum
perturbative results and tree--level relations between the lattice
parameters and the continuum ones. The lines of constant values for
the cut--off $1/a_4 \sqrt{\sigma}$ and for the scalar mass
$m_5/\sqrt{\sigma}$ must be determined non--perturbatively using
numerical simulations, and we shall see if and how they deviate from
the perturbative expectations. In particular, it would be interesting
to see if the hierarchy between the scalar mass and the string tension
still holds at stronger couplings.
%
%%%%%%%%%%%%%%%%%%%%%%%%%%%%%%%
%
\section{The phase diagram}
\label{sec:phase-diagram}
In this section, we briefly describe a further issue arising in the
study of the lattice model. Indeed, the perturbative predictions we
referred to in Sec.~\ref{sec:scale-separation} do not take into
account the rich phase structure of the lattice theory. Since it is
crucial for our purposes to simulate the theory in the correct phase,
let us first discuss the current understanding of the phase diagram of
the $\SU{2}$ pure gauge theory in five dimensions described by the
action in Eq.~\eqref{eq:aniso-lattice-action-E}.\\
The first feature, which was already investigated in the early studies
in Ref.~\cite{Creutz:1979dw}, concerns the lattice model on the line
$\beta_4 = \beta_5$, or equivalently $\gamma = 1$. This isotropic
model, where the lattice spacings are the same, $a_4 = a_5$, has a
bulk phase transition when all the dimensions are equal. This phase
transition is independent of the physical volume of the system; it is
signalled by a sudden jump of the plaquette expectation value and by a
hysteresis cycle. The bulk line separates a confined phase that is
connected to the strong coupling regime from a Coulomb--like phase
connected to the weak coupling regime. An interesting feature of the
isotropic model is that the bulk transition disappears when the
lattice size in any one dimension is decreased below a critical size,
$L_{c}$, which is the critical length of the Polyakov loop in that
direction. Below $L_c$ centre symmetry is broken. In this case the
phase transition becomes a second order one in the same universality
class of the four--dimensional Ising model: the position of the
critical point scales with the four--dimensional volume and with the
number of sites $N_5$ in the extra dimension. This has been verified
numerically for $N_5=2$ with a Binder cumulant finite--size scaling in
Ref.~\cite{deForcrand:2010be}. We have performed a scaling analysis of
the susceptibility of the Polyakov loop in the compact direction
$L_5$, and obtained compatible results. However, when the number of
points in the compact fifth dimension is increased to $N_5=4$, we
could not locate the second order phase transition before hitting
again the bulk transition; this is true up to $N_4=14$, which is the
biggest lattice we explored at $\gamma = 1$. A bigger aspect ratio
$N_4/N_5$ is probably needed at $\gamma = 1$ in order to see the
effects of the compactification (namely a thermal--like second order
phase transition), our computational resources did not allow us to
further explore this issue. A very recent study of the phase diagram
at very small anisotropies $\gamma < 1$ was presented in
Ref.~\cite{Knechtli:2010sg} and the authors claim that if any one of
the dimensions becomes smaller than a minimal lattice size $L^{\rm
  min}(\gamma)$, no sign of the bulk phase transition can be
detected. At $\gamma=1$, their simulations hint at $2 < L^{\rm min} <
6$, and the results are also supported by
Ref.~\cite{Farakos:2010ie}.\\
In the following, we are interested in the region of the parameter
space where $\gamma > 1$. Clearly, in this case % $\xi > 1$ and
$a_4 > a_5$, and hence the extra dimension can be easily
made small enough to obtain dimensional reduction as described
above. The phase diagram in this region is known at $N_5 = 4$ and $N_5
= 6$~\cite{Ejiri:2000fc}. We performed a similar study on bigger
four--dimensional lattices and obtained compatible results. We want to
stress that our aim is not to study the details of these phase
transitions; therefore we simply determined the critical lines in the
parameter space searching for the location of the peak in the
susceptibility of the compact Polyakov loop. As shown in
Fig.~\ref{fig:compare-phase-diagram} our results compare favourably to
Ref.~\cite{Ejiri:2000fc}, which also provides a cross--check of the
validity of our simulation code. \\
\FIGURE[ht]{
\epsfig{figure=FIGS/phase_diagram_cfr_ejiri_knechtli.eps,width=0.75\textwidth,clip}
\label{fig:compare-phase-diagram}
\caption{Phase diagram of the five--dimensional $\SU{2}$ pure gauge
  lattice model. In red we report the position of the bulk phase
  transition characteristic of the system with no compact
  dimensions. Both results from Ref.~\cite{Ejiri:2000fc}, and from
  Ref.~\cite{Knechtli:2010sg}, are shown. Moreover, we report the
  lines of second order phase transition at $N_5=2$ (blue), $N_5=4$
  (black) and $N_5=6$ (green). Our results are plotted with filled
  circles and come from simulations on the lattices shown in the
  legend of the plot. They are compatible with other results in the
  literature.}  }

The main feature of the phase diagram in this region is that, for
fixed $N_5$, there is a line of second order phase transition that
merges into the bulk one as the anisotropy is decreased below a
critical value $\gamma_c$. Above this $\gamma_c$, which depends on
$N_5$, the transition line separates a phase where the centre symmetry
in the extra compact direction is not broken (at smaller $\beta_5$)
from the phase where the symmetry is broken and the compact Polyakov
loop acquires a non--zero expectation value. We refer to this phase as
the dimensionally reduced one, following the terminology in
Ref.~\cite{deForcrand:2010be}. However, for $\gamma < \gamma_c$ the
bulk phase transition line separates a confined phase (at smaller
$\beta_4$) from a Coulomb--like phase extending to the weak--coupling
regime, exactly as we described in the isotropic case. This pattern of
phase transitions is shown in Fig.~\ref{fig:split-phase-diagram} using
the data already shown in Fig.~\ref{fig:compare-phase-diagram}, but
now separating the phase diagram at $N_5=4$ from the one at
$N_5=6$. Since the second order phase transition is physical, its
location changes as we change $N_5$. Note also that, at fixed $N_5$,
there is no sign of a bulk phase transition for $\gamma >
\gamma_c$. The emerging physical picture tells us that the
disappearance of the bulk phase transition happens when the
five--dimensional system compactifies; in other words, $\gamma_c$
defines a critical lattice spacing in the extra dimension $a_{5c}$
such that $L_{5c} =a_{5c} N_5$.
\FIGURE[ht]{
\epsfig{figure=FIGS/phase_diagram_splitview.eps,width=0.85\textwidth,clip}
\label{fig:split-phase-diagram}
\caption{Phase diagram of the five--dimensional $\SU{2}$ pure gauge
  lattice model in the ($\beta_4,\beta_5$) plane for different values
  of $N_5$. The data are the same as in
  Fig.~\ref{fig:compare-phase-diagram}. The bulk phase transition
  separating a confined from a Coulomb--like phase disappears for
  $\gamma > \gamma_c$ into a physical thermal--like phase
  transition. The location of this transition changes in the parameter
  space as we change $N_5$. The region we are interested in studying
  is the one labelled by $L_5 < L_{5c}$.}  }
Ref.~\cite{Ejiri:2000fc} presents estimates for $\gamma_c$, for both
$N_5=4$ and $N_5=6$, and for the renormalized anisotropy $\xi$ in
those points, so that the critical radius $R_c$ of
the extra dimension in units of the four--dimensional lattice
spacing can be computed. The data in Ref.~\cite{Ejiri:2000fc} suggest that $L_{5c}
\sim 2.25 a_4$ for $N_5=4$, and $L_{5c} \sim 2.16 a_4$ for $N_5=6$:
for extra dimensions bigger than these approximate values, the system
shows a bulk phase transition characteristic of the five--dimensional
model. The interesting region for our purposes, is at $\gamma >
\gamma_c$ and above the line of second order phase transition, where
the extra dimension is smaller than its critical value $L_5 < L_{5c}$.

%
%%%%%%%%%%%%%%%%%%%%%%%%%%%%%%%
%
\section{Lines of constant physics}
\label{sec:results}

\subsection*{Strategy of lattice simulations}
\label{sec:strategy}

Our main goal is to study whether a light scalar particle does exist
in the low--energy spectrum of the five--dimensional theory.  The
strategy of the simulations is very straightforward in principle. The
lattice model we described in Sec.~\ref{sec:lattice-setup} has four
tunable parameters: the two coupling constants $\beta_4$ and
$\beta_5$, and the number of sites $N_5$ and $N_4$. If we assume, for
the moment, that the spectrum does not depend on $N_4$ (e.g. we are in
the infinite volume limit of the lattice theory), we are left with
three parameters. Fixing the bare coupling constants dynamically
determines the two lattice spacings, whereas fixing $N_5$ determines
the length of the extra dimension. In other words, by fixing a point
in this three--dimensional bare parameter space, we are choosing a
system with a given separation of scales between the ultra--violet
cut--off $\LambdaUV$, the compactification scale $\LambdaR$ and the
scalar mass $m_5$ (all the energy scales are again expressed in units
of the string tension).\\
The three energy scales of the system can only be determined a
  posteriori by measuring physical observables with numerical Monte
Carlo simulations. \\
Let us focus first on the determination of the cut--off scale. From
Eq.~\eqref{eq:cutoff-lattice} it is clear that a measure of the
four--dimensional string tension in units of the lattice spacing
yields the separation between the low-energy scale
and the cut--off. The string tension $\sqrt{\sigma}$ in units of the
four dimensional lattice spacing $a_4$ can be extracted using
different observables. We choose to measure correlation functions of
Polyakov loops winding around the three spatial directions: the string
tension can then be extracted from the mass of the lightest state that
propagates.\\
The non--perturbative scalar mass instead can be obtained from the
ratio of two lattice observables as expressed in
Eq.~\eqref{eq:scalar-lattice}. Having obtained the string tension, we
only need to measure $m_5$ in units of the lattice spacing $a_4$.
Since it is the mass of a scalar particle, we use correlation
functions of operators that only project on the $0^{++}$
representation of the symmetry group of the cube (with positive parity
and charge) following standard spectroscopic notation. Due to the
presence of the extra dimension, different types of basis operators
can be used in the correlation functions; in particular we distinguish
those created using Polyakov loops wrapping around the compact fifth
dimension from those created using Wilson loops embedded in the three
large spatial directions. We generically refer to the first kind of
operators as the scalar ones, while the second set is referred to as
glueballs. In the following, we will focus mostly on masses extracted
from correlators of the scalar operators, but part of our analysis
will be dedicated to glueballs as well. In this respect, we greatly
improve the exploration of the scalar spectrum as first presented in
Ref.~\cite{deForcrand:2010be}. More details on the operators and on
the noise--reduction techniques we used are given in the Appendix.\\
The last scale we need to set is the compactification scale
$\LambdaR$; unfortunately, we were not able to measure a third
independent observable that could be used for this purpose. In
particular, we would need a measure of the extra dimensional lattice
spacing $a_5$ that needs to be done in the confined phase.
However this problem can be easily overcome. As we
noted in Eq.~\eqref{eq:separation-lattice}, the separation between the
cut--off and the compactification scale is determined, at leading
order, by the bare parameters of the lattice model. Therefore, knowing
$\LambdaUV$ from a measure of $a_4\sqrt{\sigma}$ at one point
$(\beta_4,\beta_5,N_5)$ is sufficient to approximately estimate
$\LambdaR$. As we already mentioned, the last step of
Eq.~\eqref{eq:separation-lattice} is only valid in the weak--coupling
limit that is reached, at fixed $\gamma$, when $\beta \rightarrow
\infty$. Although this seems like a reasonable approximation in
Ref.~\cite{deForcrand:2010be} where the values of $\beta$ are large,
we try to estimate the systematic deviation of
$\frac{\LambdaUV}{\LambdaR}$ from its tree--level value
$\frac{N_5}{\gamma}$. As we will show in the following, our
simulations are performed in a different region of the phase diagram
with respect to Ref.~\cite{deForcrand:2010be} and our $\beta$ values
are smaller.
\TABLE[htb]{
  \begin{tabular}[h]{|c|c|c|c|}
    \hline
    \multicolumn{4}{|c|}{$\xi = a\gamma^2 + b\gamma + c$} \\
    \hline \hline
    $a$ & $b$ & $c$ & $\tilde{\chi}^2$ \\
    \hline
    -- & 1.600(15) & -0.446(37) & 0.61 \\
    \hline
    -0.03(1) & 1.767(62) & -0.641(76) & 0.32 \\
    \hline
    -0.06(1) & 1.950(45) & $1-a-b$ & 0.77 \\
    \hline
  \end{tabular}
  \label{tab:fittedxi}
  \caption{Parameters of the fitted function $\xi = \xi(\gamma)$. The
    all range of data was used, from $\gamma \sim 1.224$ to $\gamma =
    4$.}
}
We expect corrections to Eq.~\eqref{eq:gamma-xi} due to
quantum fluctuations. Since the non--perturbative relation between the
bare anisotropy $\gamma$ and the renormalized one $\xi$ had already
been studied for this system, we interpolated the data available in
Ref.~\cite{Ejiri:2000fc}, in order to estimate the ratio
$\frac{\LambdaUV}{\LambdaR}$ for the points we simulated. The relation
$\xi = \xi(\gamma)$ is shown in Fig.~\ref{fig:plottedxi}.
\FIGURE[ht]{
  \epsfig{figure=FIGS/plottedxi.eps,width=0.85\textwidth,clip}
  \label{fig:plottedxi}
  \caption{The relation $\gamma \sim \xi$ is corrected by quantum
    fluctuations. We interpolated data from Ref.~\cite{Ejiri:2000fc} that
    were obtained by measuring $\xi$ non--perturbatively using ratios of
    suitable correlation functions. All the data are in the correct phase
    where $\sqrt{\sigma} \neq 0$. In the plot we show three different fits
    (with $\tilde{\chi}^2 \sim 0.5 - 0.7$) and the interpolation used to
    obtain $\xi$ in the regions where we performed the simulations (both
    for $N_5=4$ and for $N_5=6$). They all compare well and are hardly
    distinguishable.}
}
We performed three different fits of the data: a linear fit, a
quadratic one, and a quadratic fit imposing $\xi(1)=1$. The details of
the fits are summarised in Tab.~\ref{tab:fittedxi}. In practice, to
obtain $\xi$ for the points in our simulations, we used a cubic
interpolation nested inside a bootstrap procedure for the errors. The
result is again shown in Fig.~\ref{fig:plottedxi} together with the
$1$--$\sigma$ contour. The errors are such that all the lower order
fits are compatible with this interpolation. Although we only use
interpolated values of $\xi$, it must be noted that $\xi =
\xi(\gamma)$ could in principle also depend on the other bare
parameter $\beta$.
However, in the region where $\xi$ was initially measured
non--perturbatively, the value of $\xi$ is shown to be fairly
insensitive to the values of $\beta$ 
(cfr. Fig.~1 in Ref.~\cite{Ejiri:2000fc}).  This is true in particular
for the values of $\xi$ that we are going to use, namely $1.7 \leq \xi
\leq 3.0$. We performed our simulations at $\beta$ values
that are inside (or just slightly off) the region where $\xi$ was
observed to be independent of it. Hence we expect the systematic
errors of this interpolation procedure to be under control.\\
Before showing the details of the simulations and the results, let us
summarize the main steps of this study:
\begin{enumerate}
\item we fix a point in the three--dimensional parameter space
  $(\beta_4,\beta_5,N_5)$ that is in the dimensionally reduced phase;
\item on this point we measure $a_4\sqrt{\sigma}$ and $a_4m_5$ from
  correlation functions of suitable operators;
\item the measured observables determine the cut--off scale and the
  scalar mass from Eq.~\eqref{eq:cutoff-lattice} and
  Eq.~\eqref{eq:scalar-lattice};
\item we use the available data for $\xi$ to estimate the
  compactification scale using Eq.~\eqref{eq:separation-lattice} and
  the measured cut--off scale (this yields a better determination of
  the anisotropy than the one coming from 
  tree--level relation $\gamma = \xi$, and allows us to estimate the
  errors due to $\beta \neq \infty$);
\item we then move to a different point $(\beta_4,\beta_5,N_5)$ and
  repeat the procedure;
\item having done this for a certain number of points allows us to
  study the dependence of the energy scales on the bare parameters and
  to determine lines of constant physics;
\item more importantly, this allows us to study the behaviour of $m_5$ as a
  function of $\LambdaUV$ or $\LambdaR$ and to disentangle cut--off
  effects from compactification effects.
\end{enumerate}

\subsection*{Results from lattice simulations}
\label{sec:results-lattice}

\FIGURE[hb]{
  \begin{tabular}{cc}
    \epsfig{figure=FIGS/region_b4b5_n5-4.eps,width=0.47\textwidth,clip}
    &
    \epsfig{figure=FIGS/region_b4b5_n5-6.eps,width=0.47\textwidth,clip}
    \\
    {\scriptsize (a)} & {\scriptsize (b)} \\
  \end{tabular}
  \label{fig:points}
  \caption{The plots show the region of the phase diagram that we explored with
    numerical simulations, both for $N_5=4$ (a) and $N_5=6$ (b). The
    location of the second order phase transition is also shown.
    The blue squares are points where the scalar mass $a_4m_5$ was reliably extracted, whereas
    the green circles represent points where we were able to measure the string tension $a_4\sqrt{\sigma}$.}
}
We performed simulations at two different values of $N_5$ and several
different four--dimensional volumes. The smaller lattice has $N_4=10$
and $N_5=4$. This volume is also the one we used to locate the
position of the second order phase transition in the left panel of
Fig.~\ref{fig:split-phase-diagram}. On this lattice, we generated
$\mathcal{O}(800000)$ configurations and the correlators of the
interesting observables were binned over $20$ configurations after
thermalization. We chose a wide range of values for $\beta_4$ and for
$\beta_5$, starting very close to the line of second order phase
transition. In this region we expect a light scalar in units of
the lattice spacing because the scalar mass is the inverse of the correlation
length, and the latter diverges at the critical point. From the phase
structure discussed above, we also expect to find a finite string
tension. The details of the simulated points on this
volume are reported in Tab.~\ref{tab:points-n5-4}.\\
Similarly, for $N_5=6$ we simulated on lattices with $N_4=12$,
generating $\mathcal{O}(600000)$ configurations, binning the
observables over $20$ configurations. The parameters of the
simulations on this second volume are summarised in
Tab.~\ref{tab:points-n5-6}. In the tables we show both the bare
parameters that fix the location of the point in the phase diagram,
and the values of $\gamma$ and the corresponding interpolated value of
$\xi$. The size of the non--perturbative effects on the anisotropy can
be extracted from these numbers. Moreover, from the same tables, one
can compare the separation of scales in
Eq.~\eqref{eq:separation-lattice} to the naive estimate using the bare
parameters, i.e. $\frac{N_5}{\gamma}$. What we notice is that the
naive expectation is systematically larger than what is obtained by
measuring the anisotropy non--perturbatively. As a result, we were
only able to explore the following range
\begin{equation}
  \label{eq:separation-explored}
  1.7 \; \lesssim \; \frac{N_5}{\xi} = \frac{\LambdaUV}{\LambdaR} = \frac{2
    \pi R}{a_4} \; \lesssim \; 2.3
  \comma
\end{equation}
where the upper limit is close to the critical value $L_{5c}/a_4 \sim
2.16
- 2.25$ that we identified in Sec.~\ref{sec:phase-diagram}.\\
Since this is the first time that this particular region of the phase
space is explored with lattice simulations, we performed a broad scan,
aiming primarily at identifying the interesting region. As a
consequence, there are lattices for which we were unable to measure
precisely both the string tension and the scalar mass. In
Fig.~\ref{fig:points}, we show all the points reported in
Tab.~\ref{tab:points-n5-4} and in Tab.~\ref{tab:points-n5-6}, but at
the same time we identify the ones where either $a_4m_5$ or
$a_4\sqrt{\sigma}$ could not be extracted satisfactorily.\\

Our lattice data suggest that the lattice spacing $a_4$
changes dramatically in these regions of the phase space.  As shown in
Fig.~\ref{fig:points}, the string
tension $a_4\sqrt{\sigma}$ can only be measured in a small subset of
points; the points closer to the line of second order phase transition
are characterized by spatial Polyakov loops whose mass is too high for
a signal to be extracted reliably. Since the mass of the loops is
given by $N_4 a_4 \sigma$, we see that in this region the lattice
spacing $a_4$ is getting larger in units of the string tension.
Following our discussion in Sec.~\ref{sec:scale-separation}, we
regard the region close to the phase transition line as the one
characterized by a small cut--off $\LambdaUV$. In this region, there
is not a clear separation between the low--energy physics and the
cut--off, and we expect to observe large discretization errors. To
make things even more interesting, we find the scalar mass $a_4m_5$ to
be small in this same region, where $a_4$ is large. In fact, it
turns out to be very difficult to find points in the phase diagram
where both $\sqrt{\sigma}$ and $m_5$ are separated from the cut--off
scale at the same time. This results in a scalar mass $m_5 \gtrsim
\sqrt{\sigma}$ for all the points on which we were able to reliably
measure the string tension, indicating the same hierarchy expected
from perturbation theory (cfr. Fig.~\ref{fig:perturbative}). On the
other hand, a light non--perturbative scalar 
does exist very close to the second order transition line,
where $a_4m_5$ is small and $a_4\sqrt{\sigma}$ is large.\\
\FIGURE[ht]{
  \begin{tabular}{cc}
    \epsfig{figure=FIGS/b4fixed_scalar_n5-4.eps,width=0.47\textwidth,clip}
    &
    \epsfig{figure=FIGS/b5fixed_scalar_n5-4.eps,width=0.47\textwidth,clip}
    \\
    {\scriptsize (a)} & {\scriptsize (b)} \\
  \end{tabular}
  \label{fig:scalar-n5-4}
  \caption{(a) The scalar mass in units of the lattice spacing
    $a_4m_5$ as a function of $\beta_5$ and for four different values
    of $\beta_4$ at $N_5=4$. The approximate location of the critical
    region is shown by the shaded regions for the different values of
    $\beta$. (b) At fixed $\beta_5$, we show the behaviour of
    $a_4m_5$. Smaller values of $\beta_4$ are closer to the phase
    transition line, but we do not have an estimate of its location in
    this direction. Open symbols correspond to alternative fitting
    ranges in the effective mass plateaux for the scalar state (see
    Appendix). When not quoted, errors are smaller than symbols.}
}
A more quantitative statement can be made by looking at the measured
observables as functions of the bare parameters. For example, our data
allow us to study the behaviour of $a_4\sqrt{\sigma}$ at fixed value
of $\beta_4$ as we change $\beta_5$, and vice versa. The same can be
done with $a_4m_5$ and therefore with the ratio
$m_5/\sqrt{\sigma}$. In Fig.~\ref{fig:scalar-n5-4}(a) we select four
different values of $\beta_4$ and we plot the mass $a_4m_5$ obtained
from scalar operators as a function of
$\beta_5$. Fig.~\ref{fig:scalar-n5-4}(b) shows the dependence of the
scalar mass as a function of $\beta_4$ for fixed $\beta_5$. The values
of $a_4m_5$ are taken from Tab.~5 % \ref{tab:scalars-n5-4}
where we
summarise our results for $N_5=4$, whereas we report the results for
$N_5=6$ in Tab.~7% \ref{tab:scalars-n5-6}
. As we have already mentioned,
we notice that the scalar mass approaches the cut--off scale $a_4m_5
\gtrsim 1$ as we move away from the line of second order phase
transition. This happens both in the $\beta_4$ and $\beta_5$
directions. Similarly, following Ref.~\cite{deForcrand:2010be}, we can
move in the parameter space along a line of fixed $\gamma$, while
changing $\beta$. %insert mixing discussion here
We choose $\gamma \approx 1.54$ in order to obtain a separation of
scales $\LambdaUV/\LambdaR \approx 2$ after taking into account the
renormalized anisotropy. In the interval $\beta \in [ 1.71,\; 1.77 ]$,
we accurately study the low--lying spectrum of scalar particles
employing our larger set of operators with the inclusion of
glueballs. Using a variational method, detailed in the Appendix, we
studied the operator content of the different mass eigenstates in the
scalar channel. We extracted the mass of the scalar ground state and
its first excitation. The resulting masses are shown in
Fig.~\ref{fig:scalar-gamma}, where we compare the
non--perturbative scalar masses calculated via the variational ansatz
with the masses obtained solely from effective mass plateaux of scalar
operators. It is clear from the results in the plot that a variational
analysis is crucial to identify the lightest scalar state as $\beta$
is increased.\\
Further information can be obtained by studying the change in the
operator content of the scalar eigenstates as $\beta$ increases.  We
measure the normalized projection of the mass eigenstates onto each
operator used in the correlation matrix. The projection of the
extracted ground state is shown in Fig.~\ref{fig:mixing-ground}. The
plot clearly shows how the contribution of the scalar operators to the
ground state decreases as $\beta$ increases. At higher values of
$\beta$, glueball operators have a larger overlap onto the ground
state. On the other hand, we clearly see that at lower values of
$\beta$, closer to the line of second order phase transition, the
scalar state has a dominant contribution from the
extra--dimensional operators.\\
The relative mixing of the first excited state onto the operators in
the variational set is shown in Fig.~\ref{fig:mixing-exc}. The points
where the mixing is calculated are the same as in
Fig.~\ref{fig:mixing-ground}. Up to $\beta = 1.75$, the first excited
state is dominated by a projection onto the scalar operators,
suggesting an extra--dimensional nature for this particle.\\
Again, we conclude that the scalar mass becomes heavy in units of the
cut--off scale while moving away from the critical line, as shown in
Fig.~\ref{fig:scalar-gamma}. This suggests that at $\beta \gtrsim
1.77$ for $N_5=4$ the scalar particle becomes heavy; from data in
Ref.~\cite{deForcrand:2010be} taken at $1.83 \leq \beta \leq 1.91$ at
the same $N_5$ (but at $\gamma=2$) it can be shown that $m_5 \gtrsim 2
\LambdaUV$ and therefore the scalar particle cannot be considered a
low--energy degree of freedom of the theory.\\
\FIGURE[hbt]{
  \epsfig{figure=FIGS/variational_masses_cfr_scalar.eps,width=0.77\textwidth,clip}
  \label{fig:scalar-gamma}
  \caption{The scalar mass in units of the lattice spacing
    $a_4m_5$ as a function of $\beta$ for a fixed value of
    $\gamma=1.5433$. The shaded area is the approximate location of the
    second order phase transition. Open symbols refer to masses
    obtained from a variational procedure. Filled symbols are masses
    extracted from diagonal correlators of scalar operators.}
}
\FIGURE[ht]{
  \begin{tabular}{ccc}
    \epsfig{figure=FIGS/proj_ground_1.72.eps,width=0.31\textwidth,clip}
    &
    \epsfig{figure=FIGS/proj_ground_1.73.eps,width=0.31\textwidth,clip}
    &
    \epsfig{figure=FIGS/proj_ground_1.74.eps,width=0.31\textwidth,clip}
    \\
    \epsfig{figure=FIGS/proj_ground_1.75.eps,width=0.31\textwidth,clip}
    &
    \epsfig{figure=FIGS/proj_ground_1.76.eps,width=0.31\textwidth,clip}
    &
    \epsfig{figure=FIGS/proj_ground_1.77.eps,width=0.31\textwidth,clip}
    \\
  \end{tabular}
  \label{fig:mixing-ground}
  \caption{Relative projection of the ground state onto each
    of the operators in the variational set. Filled symbols
    correspond to the set of smeared scalar operators, whereas the
    open symbols refer to the smeared versions of glueball operators.}
}
\FIGURE[ht]{
  \begin{tabular}{ccc}
    \epsfig{figure=FIGS/proj_exc_1.72.eps,width=0.31\textwidth,clip}
    &
    \epsfig{figure=FIGS/proj_exc_1.73.eps,width=0.31\textwidth,clip}
    &
    \epsfig{figure=FIGS/proj_exc_1.74.eps,width=0.31\textwidth,clip}
    \\
    \epsfig{figure=FIGS/proj_exc_1.75.eps,width=0.31\textwidth,clip}
    &
    \epsfig{figure=FIGS/proj_exc_1.76.eps,width=0.31\textwidth,clip}
    &
    \epsfig{figure=FIGS/proj_exc_1.77.eps,width=0.31\textwidth,clip}
    \\
  \end{tabular}
  \label{fig:mixing-exc}
  \caption{Relative projection of the first excited state onto each
    of the operators in the variational set. Filled symbols
    correspond to the set of smeared scalar operators, whereas the
    open symbols refer to the smeared versions of glueball operators.}
}
While the scalar mass becomes smaller as we approach the critical
line, the opposite happens to the string tension. Its behaviour in
bare parameter space is best illustrated by the data at $N_5=6$. All
the points where we were able to extract the string tension
$a_4\sqrt{\sigma}$ are summarized in Tab.~\ref{tab:torelons-n5-4} for
$N_5=4$, and Tab.~\ref{tab:torelons-n5-6} for $N_5=6$. In
Fig.~\ref{fig:string-n5-6}(a) the string tension is shown at three
different values of $\beta_4$: the common feature of the data is that
the string tension increases as the critical line is
approached. As discussed above, this behaviour can be interpreted as
an increase of the lattice spacing $a_4$ in units of the physical
string tension $\sqrt{\sigma}$. A similar functional dependence of
$a_4\sqrt{\sigma}$ is shown in Fig.~\ref{fig:string-n5-6}(b), where
$\beta_5$ is fixed. At lower values of $\beta_4$, closer to the line
of phase transition, the string tension grows and it becomes very
difficult to extract a signal from our numerical simulations. We can
easily infer from the data that the string tension will decrease with
increasing $\beta$ at fixed $\gamma$, as already reported in
Ref.~\cite{deForcrand:2010be}. This behaviour is expected since $\beta
\rightarrow \infty$ is the weak--coupling limit
of the theory, and accordingly the string tension should vanish.\\
\FIGURE[ht]{
  \begin{tabular}{cc}
    \epsfig{figure=FIGS/b4fixed_string_n5-6.eps,width=0.47\textwidth,clip}
    &
    \epsfig{figure=FIGS/b5fixed_string_n5-6.eps,width=0.47\textwidth,clip}
    \\
    {\scriptsize (a)} & {\scriptsize (b)} \\
  \end{tabular}
  \label{fig:string-n5-6}
  \caption{(a) The string tension in units of the lattice spacing
    $a_4\sqrt{\sigma}$ as a function of $\beta_5$ and for three different values
    of $\beta_4$ at $N_5=6$. The approximate location of the critical
    region is shown for the different $\beta_4$. (b) At fixed
    $\beta_5$, we show the behaviour of $a_4\sqrt{\sigma}$. Smaller values of
    $\beta_4$ are closer to the phase transition line, but we do not
    have an estimate of its location in this direction. Open symbols
    correspond to alternative fitting ranges in the effective mass
    plateaux for the torelon state (see Appendix). When not quoted,
    errors are smaller than symbols.}
}
From the previous discussion we have identified the lines of constant
physics in the phase diagram at fixed $N_5$. Moreover, we find similar
features by going from $N_5=4$ to $N_5=6$. The lines of constant
cut--off $\LambdaUV$ are represented by contour lines of
$a_4\sqrt{\sigma}$. These lines start close to the line of second
order phase transition for $\gamma \sim \gamma_c$, but then move away
from it as $\gamma$ is increased. To summarize, at fixed $\gamma$, the
lowest $\beta$ corresponds to the lowest $\LambdaUV$; a larger
separation between the low--energy physics and the cut--off is found
at bigger values of $\beta$, and this is the region where the lattice
discretization starts to become irrelevant and we can safely extract
the low--energy physics from numerical simulations
(cfr. Eq.~\eqref{eq:separation-sigma-cutoff}). What we are really
interested in is the behaviour of the scalar mass $m_5$ in units of
the string tension $\sqrt{\sigma}$. By looking at the ratio of
$a_4m_5$ over $a_4\sqrt{\sigma}$, we can deduce the lines of constant
scalar mass. Unfortunately, we cannot use all the measured values of
$a_4m_5$, because we also need a measure of $a_4\sqrt{\sigma}$ on the
same point. The general pattern of these lines is again quite clear:
the lightest scalar is found closer to the second order critical line,
but it soon starts decoupling from the low--energy physics as we move
away from it. There is only a small patch of the phase space we
explored where Eq.~\eqref{eq:separation-cutoff-radius},
Eq.~\eqref{eq:separation-sigma-cutoff} and
Eq.~\eqref{eq:separation-mass-cutoff} hold simultaneously. The
lightest mass $m_5$ we measured is of order
$2\sqrt{\sigma}$.\\
Using the non--perturbative lines of constant physics, we can try to
discuss the different types of continuum limit. Our findings
can be compared with the perturbative picture reported
in Ref.~\cite{deForcrand:2010be}, bearing in mind that our results are
obtained for fixed values of $N_5$ and $N_4$ and therefore could be
affected by finite--volume effects. 
%%%%%% added after referee report %%%%%%%%%%%%
For this comparison, we shall refer in particular to Fig.~7 of
Ref.~\cite{deForcrand:2010be}.  First let us relate our choice of
parameters with the definitions in Ref.~\cite{deForcrand:2010be}: the
horizontal axis in Fig.~7, is labelled by $\beta_5$, which corresponds
to what we call $\beta$ (cfr. Eq.~\eqref{eq:beta}) in this work; the
vertical axis is labelled by $\tilde{N}_5$, which is defined as
$N_5/\gamma$. In the following we use only our parametrization,
and the reader should refer to the discussion above for any
comparison.
%%%%%%%%%%%%
At any fixed value for $\beta_4$ in the dimensionally reduced phase,
there is a lower bound for $\beta_5$, given by the location of the
critical point. By increasing $\beta_5$, we cross lines of decreasing
lattice spacing $a_4$, therefore moving towards a continuum limit,
meaning that the lattice discretization effects vanish. At the same
time we cross lines of increasing scalar mass $m_5$, which inevitably
decouples from the low--energy spectrum: the low--energy effective
theory described in this region is four--dimensional, and contains
only gauge degrees of freedom. A similar limit occurs at fixed
$\beta_5$ and increasing $\beta_4$. However, by following a line of
constant scalar mass in the phase diagram, we cross lines of different
fixed lattice spacing. In particular, moving towards smaller $\beta_4$
and bigger $\beta_5$ the lattice spacing decreases, allowing us to
reach the desired separation between the cut--off and the low--energy
physics with a constant value of the scalar mass. The low--energy
dynamics is then described by an effective four--dimensional theory
with a light adjoint scalar in the low--energy spectrum, having
started with a five--dimensional theory with only gauge degrees of
freedom. This being an effective description, it is expected to hold
only up to the energy scales given by the compactification radius, as
we already mentioned in Sec.~\ref{sec:scale-separation}. What we have
learned from our non--perturbative map of the energy scales in the
phase diagram of the lattice model is that it requires a certain
amount of fine tuning to pin down the location of a line of constant
mass and to follow it. Moreover, the behaviour of the cut--off scale
near to the line of second order phase transition
(cfr. Fig.~\ref{fig:string-n5-6}) makes it very difficult to determine
$\frac{m_5}{\sqrt{\sigma}}$ non--perturbatively, thereby limiting our
ability to reach values of the scalar mass that are smaller than the
square root of the string tension. This is an important result for
future studies in this context, and it was not anticipated before
using perturbative arguments. For example, looking at the perturbative
results in Fig.~\ref{fig:perturbative}, or equivalently at Fig.~7 in
Ref.~\cite{deForcrand:2010be}, where the line of phase transitions in
the $a_5 \rightarrow 0$ is taken into account, we note that the lines
of fixed $a_4\sqrt{\sigma}$ go straight into the critical line.  This
behaviour is not supported by our non--perturbative results: those
lines cannot cross the point where the phase transition occurs,
because $a_4\sqrt{\sigma}$ increases as we approach that point. Any
attempt to follow a line of constant scalar
mass would have to deal with this problem.\\

\subsection*{Compactification effects on the scalar mass}
\label{sec:scalar-mass}

So far we have only explored the behaviour of energy scales in the
bare parameter space. However, each point we have simulated on the phase
diagram corresponds to a precise location in the space given by the
three energy scales we are interested in, that are $\LambdaUV$,
$\LambdaR$ and $m_5$. We can therefore translate our results at
$N_5=4$ and $N_5=6$ into a common set of points
$(\LambdaUV,\LambdaR,m_5)$. This approach allows us to study $m_5$ as
a function of the other two energy scales, instead of the bare
parameters. From now on we express the energies $\LambdaUV$ and
$\LambdaR$ using their length counterpart, $a_4\sqrt{\sigma}$ and
$R\sqrt{\sigma}$ respectively. These two length scales are related to
each other by Eq.~\eqref{eq:separation-lattice} and they are both
measured non--perturbatively: the first is directly measured, whereas
the second relies on the interpolated data of $\xi$
from Ref.~\cite{Ejiri:2000fc}.\\
The range of values of $a_4\sqrt{\sigma}$ and $R\sqrt{\sigma}$ spanned
in our simulations is shown in Fig.~\ref{fig:radius-vs-string}, and the
data we used are summarized in Tab.~\ref{tab:results-n5-4} and
Tab.~\ref{tab:results-n5-6}. In the following plots, we report results
from $N_5=4$ together with the ones from $N_5=6$. When more than one
value for $a_4\sqrt{\sigma}$ or $a_4m_5$ is extracted for the same
$(\beta_4,\beta_5)$ point, we apply the following procedure: if the
values are compatible within one standard deviation, we plot the
weighted average as central value, and the weighted error as the
statistical error; we also use the spread of the results to estimate
the systematic error due to the choice of the effective mass
plateaux. If the values are not compatible, we use the average for the
central value, whereas the systematic error is chosen to comprise both
the lowest and the highest values.\\
With our available data, we can explore the behaviour of the scalar
mass $m_5$ in the following region of lattice spacing $a_4$
\begin{equation}
  \label{eq:scale-interval-a4}
  0.15 \; < \; a_4\sqrt{\sigma} \; < \; 0.40
  \comma
\end{equation}
and compactification radius $R$
\begin{equation}
  \label{eq:scale-interval-R}
  0.05 \; < \; R\sqrt{\sigma} \; < \; 0.12
  \point
\end{equation}
\FIGURE[ht]{
  \epsfig{figure=FIGS/radius_vs_string_sys.eps,width=0.77\textwidth,clip}
  \label{fig:radius-vs-string}
  \caption{The points in the phase space are mapped into the physical
    space of the energy scales of the system. In the plot we report the
    length scale corresponding to the ultra--violet cut--off, and the
    one corresponding to the compactification energy. Using both data
    from $N_5=4$ and $N_5=6$, we can span a larger region of this
    space. Some points have two different type of error bars described
    in the text: the thicker one is statistical, whereas the thinner
    is systematic.}
}
The major advantage of interpreting the data in terms of these
physical quantities
is that we can disentangle compactification effects from
cut--off effects. It is clear from Fig.~\ref{fig:radius-vs-string}
that we have points at different values of the lattice spacing, but at
the same value of the compactification radius. The scalar mass on
those particular points can therefore be studied at fixed
compactification scale and different cut--off scale. On the other
hand, we also have points at the same value of the lattice spacing,
but at different radii, which can be used to study the behaviour of
the scalar mass at fixed cut--off scale. From
Fig.~\ref{fig:radius-vs-string} we can also infer that increasing
$N_5$ would allow us to explore a wider range of cut--off values for
fixed
compactification scale.\\
Our main goal is to clarify the validity of the result in
Eq.~\eqref{eq:one-loop-mass} where the perturbative scalar mass is
expected to depend strongly on the compactification scale. In our
lattice model we would like to see if there are leading cut--off
corrections to this expected behaviour when we look at the
non--perturbatively measured scalar mass. The simplest way of looking
for these corrections is to study the dependence of the scalar mass on
the lattice spacing. However, our values for the lattice spacing
usually correspond to different values of the compactification
radius. It is clear from this discussion that the study in
Ref.~\cite{deForcrand:2010be} cannot give any hints about
Eq.~\eqref{eq:one-loop-mass}: the lattice spacing always changes
together with the compactification radius, because their ratio is
forced to be constant. Nothing can be said about the dependence of
$m_5$ at fixed compactification scale nor at fixed cut--off scale from
the results of these earlier studies. Using our data, we can plot
$m_5$ as a function of $a_4$ and separately as a function of $R$. The
plots are shown in Fig.~\ref{fig:scalar}: Fig.~\ref{fig:scalar}(a)
shows the scalar mass dependence on the lattice spacing in the range
defined in Eq.~\eqref{eq:scale-interval-a4}, whereas
Fig.~\ref{fig:scalar}(b) shows its behaviour as a function of the
compactification radius in the range of
Eq.~\eqref{eq:scale-interval-R}. The observed range for the scalar
mass in units of the string tension is
\begin{equation}
  \label{eq:m5sigma-interval}
  2 \; < \; \frac{m_5}{\sqrt{\sigma}} \; < \; 10
  \point
\end{equation}
The important thing to notice in this analysis is how the scalar mass
changes between the two plots. While some of the points are
insensitive to the two different choices of variables, it is striking
to see points with the same mass but far away in
Fig.~\ref{fig:scalar}(a) fall on top of each other once expressed in
terms of the compactification radius in Fig.~\ref{fig:scalar}(b). If
Eq.~\eqref{eq:one-loop-mass} holds, then the combination $m_5R$ should
be independent of $R$ at leading order, while retaining any dependence
on the cut--off $a_4$. In order to separate the scalar particle from
the Kaluza--Klein modes, this variable should be less than one as we
stated in Eq.~\eqref{eq:separation-mass-cutoff}. In 
Fig.~\ref{fig:m5R-vs-string} we plot $m_5R$ as a function of
$a_4\sqrt{\sigma}$. The data show a scalar mass in units of the
compactification radius in the range
\begin{equation}
  \label{eq:m5R-interval}
  0.2 \; < \; m_5R \; < \; 0.5
  \point
\end{equation}
Such range is smaller than the one spanned by $m_5/\sqrt{\sigma}$ by
a factor of $2$ for the same interval of lattice
spacings. This evidence support the observation that the dependence on
$a_4\sqrt{\sigma}$ is milder than the one shown in
Fig.~\ref{fig:scalar}(a) and it is compatible with the perturbative
expectation in Eq.~\eqref{eq:one-loop-mass}. The product $m_5R$ does
not show any sign of quadratic divergences as the lattice spacing is
reduced. However, we must recall that all the simulations were
performed on a fixed value of $N_4$, therefore the points at the
smallest values of $a_4\sqrt{\sigma}$ are the ones on the smallest
physical volumes and finite--size effects could be present. On the
other hand, large values of $a_4\sqrt{\sigma}$ point in the direction
of larger discretization effects.
\FIGURE[ht]{
  \begin{tabular}{cc}
    \epsfig{figure=FIGS/scalar_vs_string_sys.eps,width=0.47\textwidth,clip}
    &
    \epsfig{figure=FIGS/scalar_vs_radius_sys.eps,width=0.47\textwidth,clip}
    \\
    {\scriptsize (a)} & {\scriptsize (b)} \\
  \end{tabular}
  \label{fig:scalar}
  \caption{(a) The scalar mass as a function of the lattice spacing
    for $N_5=4$ and $N_5=6$; (b) The same scalar mass is also shown as
    a function of the compactification radius. Points at the same
    value of the radius have the same mass within the statistical
    error. Each of the point in (a) corresponds to a point in
    (b). Systematic errors due to the choice of the effective mass
    plateaux are reported for some points using thinner error bars.}
}
\FIGURE[hbt]{
  \epsfig{figure=FIGS/scalarR_vs_string_sys.eps,width=0.77\textwidth,clip}
  \label{fig:m5R-vs-string}
  \caption{The scalar mass in units of the compactification radius
    $m_5R$ is shown to be mildly dependent on the lattice spacing
    $a_4\sqrt{\sigma}$. The grey band $0.3 \leq m_5R \leq 0.4$
    includes all points within two standard deviations and it is drawn
    to guide the eye. Also in this case, systematic errors are shown
    with thinner error bars, whereas the thicker ones represent
    statistical standard deviations.}
}

%
%%%%%%%%%%%%%%%%%%%%%%%%%%%%%%%
%
\section{Conclusions}
\label{sec:conclusions}
Lattice theories in more than four dimensions prove to be very
interesting. They provide a sensible and well--defined regularization
of non--renormalizable gauge theories that can be used as UV
completions to calculate phenomenologically interesting
quantities.\\
In this work we presented a non--perturbative study of pure SU($2$)
gauge theory in five dimensions. The system was discretized on
anisotropic lattices and we investigated the so--called dimensionally
reduced phase, where a light scalar particle is expected
in perturbation theory due to
compactification of the extra dimension.\\
If the scales of the theory are properly separated, we expect the
low--energy dynamics of this theory to be described by a four--dimensional
gauge theory coupled to a scalar field. We have measured the mass of
the non--perturbative scalar states in a specific region of the bare parameter space,
where we expect to find the desired separation between physical
scales. We have also determined numerically the four--dimensional
lattice spacing in units of the string tension. This allowed us to
describe the lines of constant scalar mass and constant ultra--violet
cut--off as they arise non--perturbatively.
%%%%%%% added after referee report %%%%%%%
The scale separation of Eq.~\eqref{eq:separation-mass-cutoff} is
obtained from simulations in a given region of the phase diagram, and
is shown in Fig.~\ref{fig:scalar} and Fig.~\ref{fig:m5R-vs-string}.
%%%%%%%%%%%%%
The final picture seems to
confirm the observation in Ref.~\cite{deForcrand:2010be} about the
possibility of effectively describing a four--dimensional Yang--Mills
theory with a scalar adjoint particle in the continuum limit. While
that observation was based entirely on perturbative results, our
numerical simulations show how this continuum limit could be actually
reached by following lines of constant scalar mass in the parameter
space of the model. As we described in Sec.~\ref{sec:results}, this is
not a straightforward procedure and it definitely requires some sort
of fine tuning.
Even though the search for a light scalar requires fine tuning in this
simple model, we have shown that its mass is only very mildly affected
by the ultra--violet cut--off, whereas it strongly depends on the
radius of the compactified extra dimension. This is entirely
compatible with the perturbative result of
Eq.~\eqref{eq:one-loop-mass} and it is the first non--perturbative
evidence that the mass of scalar particles coming from a
compactification mechanism does not have a quadratic dependence on the
cut--off. We need to bear in mind that our results are obtained at
finite lattice spacings, both $a_4$ and $a_5$, and at finite
volume. Therefore, it would be ideal to extend our study on larger
lattices, with $N_4 > 12$ and $N_5 > 6$, in order to reduce the
systematics of finite--size effects and discretization effects.\\
We can consider our study as a starting point for exploring in more
details the realization of gauge theories with light scalar particles
in the framework of dimensional reduction. In particular, it would be
interesting to compare the non--perturbative spectrum of the
five--dimensional model obtained in this study with the
non--perturbative spectrum of a four--dimensional gauge theory with a
scalar degree of freedom. A similar comparison has been carried on
between four-- and three--dimensional gauge
theories~\cite{Hart:1999dj}, where the 
super--renormalizability of the latter helped in the definition of
physical observables. Another interesting future
extension of this work could be the inclusion of
fermionic degrees of freedom, which are expected to
further reduce the scalar mass~\cite{DelDebbio:2008hb}, at least at a
one--loop level. As a consequence, they might allow the description of
theories with extended supersymmetry on the lattice without fine
tuning the scalar mass.

\acknowledgments It is a pleasure to thank Rodolfo Russo and Richard
Kenway for discussions and comments on this manuscript. This work has
made use of the resources provided by the Edinburgh Compute and Data
Facility (ECDF). ( http://www.ecdf.ed.ac.uk/). The ECDF is partially
supported by the eDIKT initiative (http://www.edikt.org.uk).\\
ER is supported by a SUPA prize studentship. ER also aknowledges
hospitality and support from the INFN, Laboratori Nazionali di
Frascati, during the final stage of this work. 

%
%%%%%%%%%%%%%%%%%%%%%%%%%%%%%%%
%
\section*{Appendix}
\label{sec:appendix}

\subsection*{Extracting the string tension and the scalar mass}
\label{sec:operators}

The string tension and the mass of the ground state in the scalar
channel have been
measured at different points $(\beta_4,\beta_5,N_4,N_5)$ in the bare
parameter space. We use standard lattice spectroscopic techniques and
we extract
the masses from $2$--point functions of suitable lattice operators
coupling to the states of interest and correlated in the time
direction (which is taken to be one of the $4$ directions with $N_4$
lattice sites). The correlators are then averaged over the $N_5$ slices in
the extra dimension.\\
To extract the string tension we use Polyakov loop operators $L_i$
winding around the $3$ spatial dimensions ($i\in \{1,2,3\}$). These
operators are non--local and have a non--zero charge under the centre
symmetry group. They couple to torelon states whose mass grows
linearly with the size of the lattice. The string tension is the
coefficent of this linear dependence; this procedure yields the right
string tension if the open--close duality between Wilson loops and
Polyakov loops holds.
More specifically, the mass of the torelon states are related to the
string tension as follows:
\begin{equation}
  \label{eq:torellon-mass}
  a_4m_{\rm tor}(N_4) \; = \; a_4^2\sigma N_4 - \frac{\pi (D-2)}{6 N_4}
  \comma
\end{equation}
where $D$ is the number of spacetime dimensions.
From the above relation Eq.~\eqref{eq:torellon-mass} we can extract the string tension as
\begin{equation}
  \label{eq:string-tension}
  a_4^2 \sigma \; = \; \frac{a_4m_{\rm tor}(N_4)}{N_4} + \frac{\pi (D-2)}{6 N_4^2} 
  \comma
\end{equation}
and we set $D=4$ in our analysis.\\
The systematic error in extracting the string tension using
Eq.~\eqref{eq:string-tension} is known to be small for long
Polyakov loops, i.e. loops such that $N_4 a_4 \sqrt{\sigma}
>3$. Unfortunately, measures of Polyakov loop operators are
difficult because of the poor signal--to--noise ratio; specific
techniques are usually needed in order to enhance the signal, and
obtain statistically accurate results. In this work, we use an improved diagonal
spatial smearing with a further step of blocking as first described
in Ref.~\cite{Lucini:2004my}. The set of parameters used here is the
same as in Ref.~\cite{Lucini:2004my}, namely
$(p_a,p_d)=(0.40,0.16)$ (cfr. Fig.~\ref{fig:smear-path}).
\FIGURE[b]{
  \epsfig{file=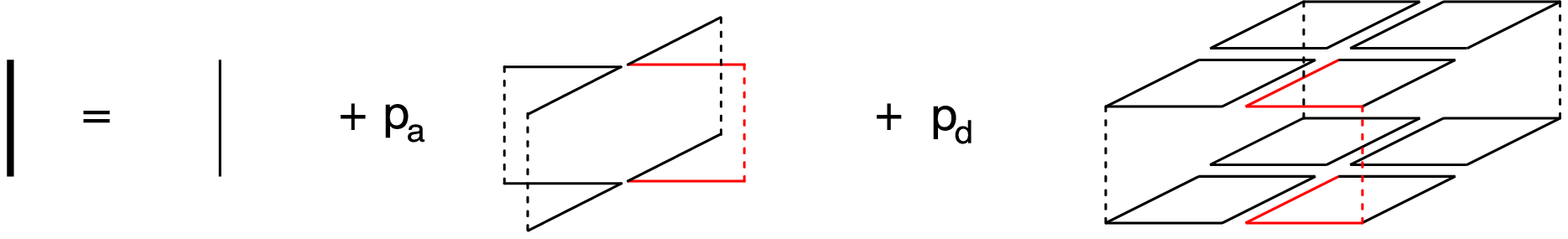,width=0.61\textwidth,clip}
  \label{fig:smear-path}
  \caption{Example of smearing of a single link. Sum of orthogonal
    staples and diagonal staples are weighted with two independent
    parameters: $p_a$ and $p_d$. The figure is taken
    from~\cite{Lucini:2010nv}.}
}
The diagonal correlators of spatial Polyakov loops at different
blocking levels are analysed using a single--state hyperbolic cosine
fit to extract the effective mass, and jackknife bins are used to
estimate the statistical errors. For all the points where we measure
the string tension, the best projection onto the ground state is
obtained at the maximum blocking level.  This was confirmed using a
variational procedure on the set of operators including all the
different blocking levels. For example, in
Fig.~\ref{fig:variational-comparison} we show the comparison between
the mass extracted from diagonal correlators of the operator at the
highest level of blocking and the one coming from the variational
procedure. The comparison was done on a lattice with a longer temporal
direction $L_t = 2 L_4$ and using the same fitting window for the
effective mass plateaux in both cases.\\
On the points $(\beta_4,\beta_5,N_4,N_5)$ used for the measurements,
we extracted the effective mass plateaux for the spatial Polyakov
loops only at large temporal distances.  The smearing and blocking
algorithm allows for the extraction of a better signal, even though
the parameters $(p_a,p_d)$ are not optimized for the broad range of
lattice spacings $a_4$ explored in this work.  In many cases, we still
have small overlaps with the ground state, and consequently the
single--state behaviour of the correlator can only be extracted at
large temporal distances. An example of such cases is shown in
Fig.~\ref{fig:tor-plateaux}(a), whereas in
Fig.~\ref{fig:tor-plateaux}(b) we show one
of the points where the plateaux is reached already at $t/a_4=2$.\\
A summary of all the torelon masses and their corresponding string
tensions is reported in Tab.~\ref{tab:torelons-n5-4}, and
Tab.~\ref{tab:torelons-n5-6}. The fitting range for the effective mass
plateaux is also shown in the tables. Moreover, since the length of
the Polyakov loops in lattice units is different between the $N_5=6$
lattices and the $N_5=4$ ones, we also report the physical size
$L_4\sqrt{\sigma}$. As mentioned above, finite--size effects can be
kept under control when $L_4\sqrt{\sigma}$ is large; in other words we
would like our physical lattice volume to be much larger than the
typical correlation length of the system, given by the
inverse of the string tension.\\
\FIGURE[ht]{
  \epsfig{file=FIGS/plateaux_0.88_3.60_cfrvar.eps,width=0.77\textwidth,clip}
  \label{fig:variational-comparison}
  \caption{Comparison between the ground state effective mass
    extracted from the diagonal correlator of the highest blocking
    level (black points) and the one extracted from the variational
    procedure (red circles). The operators were
    $L_4=12a_4$ spatial Polyakov loops at $4$ different blocking levels,
    and their correlator was measured along a $L_t=2L_4=24a_4$
    temporal distance. The correlator was averaged over the $N_5=6$
    extra dimension slices.}
}
\FIGURE[ht]{
  \begin{tabular}{cc}
      \epsfig{file=FIGS/plateaux_0.90_3.45.eps,width=0.47\textwidth,clip}
      &
      \epsfig{file=FIGS/plateaux_0.86_3.70.eps,width=0.47\textwidth,clip}
      \\
    {\scriptsize (a)} & {\scriptsize (b)} \\
  \end{tabular}
  \label{fig:tor-plateaux}
  \caption{Example of plateaux of torelon effective masses. (a) The
    final mass comes from a weighted fit of the points in the plateaux
    that is reached only at $t/a_4=3$ due to the small overlap of the
    operator onto the ground state. (b) At lower masses the plateaux
    is longer and the signal is extracted more reliably.}
}
\FIGURE[ht]{
  \begin{tabular}{cc}
    \epsfig{file=FIGS/plateaux_0.89_3.55_scalar.eps,width=0.47\textwidth,clip}
    &
    \epsfig{file=FIGS/plateaux_0.90_3.45_scalar.eps,width=0.47\textwidth,clip}
    \\
    {\scriptsize (a)} & {\scriptsize (b)} \\
  \end{tabular}
  \label{fig:scal-plateaux}
  \caption{Example of plateaux for one of the highest scalar
    masses (a) and one of the lowest (b). (a) The operator overlaps
    poorly on the ground state and the 
    plateaux is reached at large temporal distances. In this case we
    tried to estimate the systematic error on the fitting range, by
    choosing two different fitting windows. (b) The plateaux is
    reached already at $t/a_4=2$.}
}
For the mass of the static scalar mode, we use compact
Polyakov loop operators, that is gauge--invariant combinations of
Polyakov loops winding around the extra fifth dimension. Such
operators transform as scalars under the cubic symmetry group and they
only carry a site index in the four--dimensional subspace.
In particular, we choose two different combinations
\begin{equation}
  \label{eq:scalar-1}
  \OC_1(t) \; = \; \sum_x \Tr[L_5(x,t)] \; ; \qquad 
  L_5(x,t)=\prod_{j=1}^{N_5} \UC_5(x+ja_5\hat{5},t)
  \comma
\end{equation}
and
\begin{equation}
  \label{eq:scalar-2}
  \OC_2(t) \; = \; \sum_x \Tr[\phi(x,t) \phi^\dag(x,t)] \; ; \qquad
  \phi(x,t)=\frac{L_5-L_5^\dag}{2}
  \point
\end{equation}
The sum $\sum_x$ is an average over the spatial volume in order to
obtain zero--momentum operators on a fixed timeslice $t$. We average
the correlators over the
extra--dimensional coordinate, as in the previous case.\\
The first operator $\OC_1$ is the same one used in
Ref.~\cite{deForcrand:2010be}.  For the operator in
Eq.~\eqref{eq:scalar-2} it is possible to apply a smearing procedure
following the one introduced in Ref.~\cite{Irges:2006hg} for a scalar
Higgs field. The operator $\phi$ is replaced by a smeared version that
consists of a gauge--invariant combination of parallel transporters in
the
three--dimensional spatial subspace.\\
For this observable, we found the lowest smearing level of $\OC_2$ to
have the largest projection onto the ground state. The effective
masses extracted from $\OC_1$ and the lowest smearing level of $\OC_2$
are always compatible. In some cases, usually at very low masses
$a_4m_5$, the smeared operators show better plateaux, but we have not
yet studied their projection onto the ground state with a more
systematic variational procedure. At this stage we have not
implemented more efficient noise--reduction techniques, such as a
better choice for the smearing parameters, a multi--level
algorithm~\cite{Luscher:2001up}, or a multi--hit procedure. As
a consequence, we obtain plateaux like the ones shown in
Fig.~\ref{fig:scal-plateaux}(a). The mass is extracted from a weighted
fit of three, or sometimes even two, points at very large temporal
distance, where the signal--to--noise ratio is quite small. Clearly
there also points where the scalar mass is small, and the effective
mass plateaux is well behaved. An example can be found in
Fig.~\ref{fig:scal-plateaux}(b). We summarize the scalar masses
$a_4m_5$, and the fitting windows for the plateaux in
Tab.~5% \ref{tab:scalars-n5-4}
, and Tab.~7% \ref{tab:scalars-n5-6}
.\\
\FIGURE[hb]{
  \epsfig{file=FIGS/volume_effects_V12_V16.eps,width=0.77\textwidth,clip}
  \label{fig:volume-effects}
  \caption{For three different points reported on the $x$ axis, we
    show the string tension and the scalar mass in units of the
    cut--off length. Two volumes are compared and sizable finite--size
    effects can be ruled out.}
}
An attempt to estimate the finite--volume effects on the observables
$a_4\sqrt{\sigma}$ and $a_4m_5$ has been performed at $N_5=6$. For
three different points in the phase space $(\beta_4,\beta_5)$, we
simulate two different four--dimensional lattice sizes, $N_4=12$ and
$N_4=16$. The three points have a very similar string tension at
$N_4=12$, but on that volume $L_4\sqrt{\sigma}$ turns out to be
smaller than $3$. The results are summarized in
Tab.~\ref{tab:volume-effects}. The string tension and the scalar mass
are not affected by the change of four--dimensional volume. In
Fig.~\ref{fig:volume-effects} the volume dependence is shown for both
the observables. The larger statistical error for $a_4\sqrt{\sigma}$
on the largest volume is due to the larger torelon mass at $N_4=16$
(the number of configurations is the same for both volumes).\\
\TABLE[t]{
  \begin{tabular}[h]{c|c|c||c|c|}
    \cline{2-5}
    &\multicolumn{2}{|c||}{$a_4\sqrt{\sigma}$}&\multicolumn{2}{|c|}{$a_4m_5$}\\
    \hline
    $N_4$ & 12 & 16 & 12 & 16 \\
    \hline
    (0.845,3.80) & 0.2358(11) & 0.2364(25) & 1.05(31) & 1.44(9)\\
    (0.870,3.65) & 0.2358(11) & 0.2359(27) & 0.998(93) & 1.08(26)\\
    (0.870,3.65) & 0.2383(12) & 0.2392(25) & 0.777(78) & 0.956(99)\\
    \hline
  \end{tabular}
  \label{tab:volume-effects}
  \caption{Comparison between the observables on two different
    four--dimensional volumes and fixed $N_5=6$. The string tensions
    are independent of $N_4$ and the scalar masses are compatible
    within one standard deviation.}
}
\subsection*{Mixing with scalar glueball states}
\label{sec:glueballs}

Since we are studying a strongly coupled Yang--Mills theory, the
low--energy dynamics could be affected by the presence of
non--perturbative states, such as glueballs.
It is well known that the lightest glueball state appears in
the scalar channel. However, this is the same symmetry channel where
we perturbatively expect to see a light particle due to the
compactification mechanism. It is therefore mandatory to check whether
these two states mix in order to shed light on the non--perturbative
fate of Eq.~\eqref{eq:one-loop-mass}.\\
Lattice calculations of glueball masses suffer from the aforementioned problems
in relation to torelon masses: to obtain an accurate
estimate of the mass from correlation functions, one needs to adopt
noise--reduction techniques. We used a combination of the improved
diagonal smearing described in Fig.~\ref{fig:smear-path} and a
variational ansatz. We used three different spatially shaped Wilson loops
in order to construct glueball operators. This procedure has been very
succesfull in extracting highly accurate glueball
masses in three and four--dimensional $\SU{N}$ gauge
theories~\cite{Lucini:2004my,Lucini:2010nv,Morningstar:1999rf}.\\
To create operators coupling to glueball states in four dimensions, we
use the four--links plaquette, the six--links rectangular
plaquette and the six--links chair shown in
Fig.~\ref{fig:glue-path}. Symmetrized combinations of these operators
projecting only onto the scalar representation of the three--dimensional
cubic symmetry group are then correlated together with operators in
Eq.~\eqref{eq:scalar-1} and Eq.~\eqref{eq:scalar-2}; we
refer to these scalar glueball operators as $\OC_a$, $\OC_b$ and
$\OC_c$ built starting respectively from the path a), b) and c) in
Fig.~\ref{fig:glue-path} (we always use zero--momentum
projections).\\
\FIGURE[b]{
  \epsfig{file=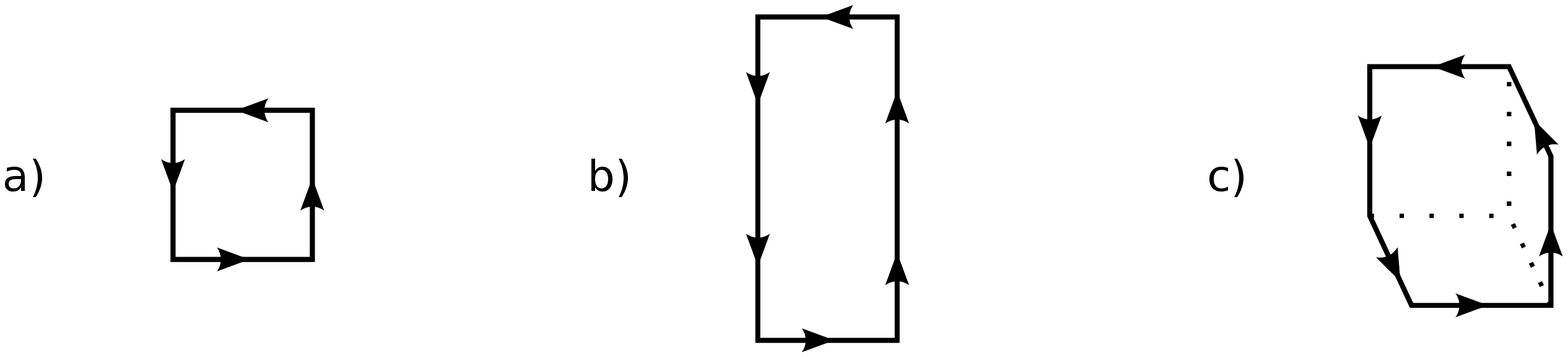,width=0.61\textwidth,clip}
  \label{fig:glue-path}
  \caption{Wilson loops used in the construction of glueball operators
    in the scalar channel. Each of these three operators is smeared
    according to Fig.~\ref{fig:smear-path} in order to construct a larger
    variational ansatz.}
}
We expect that the operators in Fig.~\ref{fig:glue-path} will couple
mainly onto glueball states as they are built entirely from links in
the three--dimensional spatial subspace of the lattice. On the other
hand, we suggest that the operators in Eq.~\eqref{eq:scalar-1} and
Eq.~\eqref{eq:scalar-2} will couple mainly with states of
extra--dimensional nature because they are
built from links in the extra direction. Inevitably, due to the
non--perturbative nature of the theory, the masses of the scalar
state extracted from correlators of the latter type of operators could
be affected by non--negligible mixing with glueball states. We studied
the contribution of this mixing, and the results are reported and explained in
Sec.~\ref{sec:results}.\\
To estimate the mixing in the scalar spectrum of the lattice theory, we used
the following procedure:
\begin{itemize}
\item compute the full correlation matrix $C_{\alpha \beta}(t)$, where the
  lower indices run over the scalar operators of the following type:
  $\OC_2$, $\OC_a$, $\OC_b$, $\OC_c$
\item employ a variational procedure to find a linear combination of
  the correlated operators such that the propagating state is the
  lightest (or apply the procedure on the orthogonal space to get the
  excited ones)
\item decompose the approximate mass eigenstates obtained from the
  previous step into their 
  projections onto the basis operators $\OC_2$, $\OC_a$, $\OC_b$,
  $\OC_c$.
\end{itemize}
The last step of this variational analysis gives us informations about
the nature of the propagating state. If the main projection is onto
glueball operators $\OC_a$, $\OC_b$ and $\OC_c$, the mass extracted is
likely to be associated to a glueball state rather than a scalar of
extra--dimensional origin. At the same time, a projection onto $\OC_2$
of more than $50\%$ indicates that the state investigated is probably
a scalar coming from the compactification mechanism.\\ 
Due to the large computational cost, we measured the full correlation
matrix $C_{\alpha \beta}(t)$ only on a subset of the
points reported in Tab.~\ref{tab:points-n5-4}: we choose points at
fixed $\gamma \approx 1.54$ and we investigate how the mixing of the
extracted states changes as we increase $\beta$, moving away from the
line of second order phase transition. On these points, the masses
extracted using only correlators of $\OC_1$ and $\OC_2$, as described
in the previous section, has been shown in
Fig.~\ref{fig:scalar-gamma}. In order to extract a more reliable
plateau, we increased the number of lattice points in the temporal
direction $L_t=2 L_4$; this allowed us to follow the plateaux of the
effective mass for a wider range of temporal distances, usually
corresponding to a fitting range $t_{\rm min}-t_{\rm max} = [3-7]$
in units of the lattice spacing $a_4$ (cfr. for example the fitting
ranges of Tab.~5). The results for the spectrum of the theory in the
scalar channel at $\gamma \approx 1.54$ is summarized in
Fig.~\ref{fig:scalar-gamma}. An example of the effective mass plateax
for the ground state and its low--energy excitations is also shown in
Fig.~\ref{fig:plateaux-scalar} for two values of $\beta$; we compare
the results of the variational procedure, with the results obtained
from diagonal correlators of pure scalar operators.\\
\FIGURE[ht]{
  \begin{tabular}{cc}
    \epsfig{file=FIGS/eff_mass_1.72.eps,width=0.47\textwidth,clip}
    &
    \epsfig{file=FIGS/eff_mass_1.76.eps,width=0.47\textwidth,clip}
    \\
    {\scriptsize (a)} & {\scriptsize (b)} \\
  \end{tabular}
  \label{fig:plateaux-scalar}
  \caption{Example of effective mass plateaux for two different values
    of $\beta$. (a) At $\beta = 1.72$ the mass of the low--lying
    scalar state obtained from the variational ansatz is compatible
    with the one we measured using only scalar operators. (b) At
    $\beta = 1.76$ the scalar operator yelds a mass which is
    compatible with the second excitations of the scalar spectrum.}
}
In the range of parameters explored with the full variational ansatz,
we notice that the relative mixing of the scalar states
in the spectrum with the different operators in the correlator matrix
changes with $\beta$. The mixing of the extracted ground state
is shown in Fig.~\ref{fig:mixing-ground}. The relative projections for
both sets of operators are shown for different values of $\beta$. The
plot clearly shows the contribution of the operator $\OC_2$ to the ground
state of the scalar channel decreasing as $\beta$ increases. We
recall here that increasing $\beta$ at fixed anisotropy corresponds to
going towards the weak--coupling limit. This is the same limit taken
in the simulations of Ref.~\cite{deForcrand:2010be}, where it has been
shown how the mass extracted from correlators of our $\OC_1$ diverges
and decouples from the low--energy spectrum. It is therefore not
surprising that the lightest glueballs become relevant to the dynamics
of the
theory in this region of the parameter space. On the other hand, we
clearly see that at lower values of $\beta$, closer to the line of
second order phase transition, the scalar state has a dominant
contribution from the extra--dimensional operator $\OC_2$ and
it increases as we lower the values of $\beta$.\\
Another interesting mixing we looked at is shown in
Fig.~\ref{fig:mixing-exc}. The plots show the relative mixing of the
first excited state onto the operators in the variational set. The
points where the mixing is calculated are the same as in
Fig.~\ref{fig:mixing-ground}. Up to $\beta = 1.75$, the first excited
state is dominated by a projection onto the scalar operator $\OC_2$,
suggesting an extra--dimensional nature for this particle. What it is
not shown is that at $\beta = 1.76$, we find the second excited state
to project mostly onto $\OC_2$ (cfr. Fig.\ref{fig:plateaux-scalar}(b)).\\

%
%%%%%%%% bibliography %%%%%%
%
%% \clearpage
\bibliographystyle{JHEP}
\bibliography{light_scalar_5D}

\providecommand{\href}[2]{#2}\begingroup\raggedright\begin{thebibliography}{10}

\bibitem{Kaluza:1921tu}
T.~Kaluza, {\it {On the Problem of Unity in Physics}},  {\em
  Sitzungsber.Preuss.Akad.Wiss.Berlin (Math.Phys.)} {\bf 1921} (1921) 966--972.

\bibitem{Klein:1926tv}
O.~Klein, {\it {Quantum Theory and Five-Dimensional Theory of Relativity. (In
  German and English)}},  {\em Z.Phys.} {\bf 37} (1926) 895--906.

\bibitem{Hosotani:1983xw}
Y.~Hosotani, {\it {Dynamical Mass Generation by Compact Extra Dimensions}},
  {\em Phys. Lett.} {\bf B126} (1983) 309.

\bibitem{Hatanaka:1998yp}
H.~Hatanaka, T.~Inami, and C.~S. Lim, {\it {The gauge hierarchy problem and
  higher dimensional gauge theories}},  {\em Mod. Phys. Lett.} {\bf A13} (1998)
  2601--2612, [\href{http://arxiv.org/abs/hep-th/9805067}{{\tt
  hep-th/9805067}}].

\bibitem{Cheng:2002iz}
H.-C. Cheng, K.~T. Matchev, and M.~Schmaltz, {\it {Radiative corrections to
  Kaluza-Klein masses}},  {\em Phys. Rev.} {\bf D66} (2002) 036005,
  [\href{http://arxiv.org/abs/hep-ph/0204342}{{\tt hep-ph/0204342}}].

\bibitem{vonGersdorff:2002as}
G.~von Gersdorff, N.~Irges, and M.~Quiros, {\it {Bulk and brane radiative
  effects in gauge theories on orbifolds}},  {\em Nucl. Phys.} {\bf B635}
  (2002) 127--157, [\href{http://arxiv.org/abs/hep-th/0204223}{{\tt
  hep-th/0204223}}].

\bibitem{Hosotani:2005fk}
Y.~Hosotani, {\it {Dynamical gauge symmetry breaking by Wilson lines in the
  electroweak theory}},  \href{http://arxiv.org/abs/hep-ph/0504272}{{\tt
  hep-ph/0504272}}.

\bibitem{Hosotani:2007kn}
Y.~Hosotani, N.~Maru, K.~Takenaga, and T.~Yamashita, {\it {Two loop finiteness
  of Higgs mass and potential in the gauge-Higgs unification}},  {\em Prog.
  Theor. Phys.} {\bf 118} (2007) 1053--1068,
  [\href{http://arxiv.org/abs/0709.2844}{{\tt arXiv:0709.2844}}].

\bibitem{DelDebbio:2008hb}
L.~Del~Debbio, E.~Kerrane, and R.~Russo, {\it {Mass corrections in string
  theory and lattice field theory}},  {\em Phys. Rev.} {\bf D80} (2009) 025003,
  [\href{http://arxiv.org/abs/0812.3129}{{\tt arXiv:0812.3129}}].

\bibitem{Antoniadis:1990ew}
I.~Antoniadis, {\it {A Possible new dimension at a few TeV}},  {\em Phys.Lett.}
  {\bf B246} (1990) 377--384.

\bibitem{Ejiri:2000fc}
S.~Ejiri, J.~Kubo, and M.~Murata, {\it {A study on the nonperturbative
  existence of Yang-Mills theories with large extra dimensions}},  {\em Phys.
  Rev.} {\bf D62} (2000) 105025,
  [\href{http://arxiv.org/abs/hep-ph/0006217}{{\tt hep-ph/0006217}}].

\bibitem{Ejiri:2002ww}
S.~Ejiri, S.~Fujimoto, and J.~Kubo, {\it {Scaling laws and effective dimension
  in lattice SU(2) Yang-Mills theory with a compactified extra dimension}},
  {\em Phys.Rev.} {\bf D66} (2002) 036002,
  [\href{http://arxiv.org/abs/hep-lat/0204022}{{\tt hep-lat/0204022}}].

\bibitem{deForcrand:2010be}
P.~de~Forcrand, A.~Kurkela, and M.~Panero, {\it {The phase diagram of
  Yang-Mills theory with a compact extra dimension}},  {\em JHEP} {\bf 06}
  (2010) 050, [\href{http://arxiv.org/abs/1003.4643}{{\tt arXiv:1003.4643}}].

\bibitem{Knechtli:2010sg}
F.~Knechtli, M.~Luz, and A.~Rago, {\it {On the phase structure of
  five-dimensional SU(2) gauge theories with anisotropic couplings}},  {\em
  Nucl.Phys.} {\bf B856} (2012) 74--94,
  [\href{http://arxiv.org/abs/1110.4210}{{\tt arXiv:1110.4210}}].

\bibitem{Farakos:2010ie}
K.~Farakos and S.~Vrentzos, {\it {Exploration of the phase diagram of 5d
  anisotropic SU(2) gauge theory}},  \href{http://arxiv.org/abs/1007.4442}{{\tt
  arXiv:1007.4442}}.

\bibitem{Creutz:1979dw}
M.~Creutz, {\it {Confinement and the Critical Dimensionality of Space- Time}},
  {\em Phys. Rev. Lett.} {\bf 43} (1979) 553--556.

\bibitem{Hart:1999dj}
A.~Hart and O.~Philipsen, {\it {The spectrum of the three-dimensional adjoint
  Higgs model and hot SU(2) gauge theory}},  {\em Nucl. Phys.} {\bf B572}
  (2000) 243--265, [\href{http://arxiv.org/abs/hep-lat/9908041}{{\tt
  hep-lat/9908041}}].

\bibitem{Kurkela:2009dg}
A.~Kurkela, P.~de~Forcrand, and M.~Panero, {\it {Dimensional reduction and the
  phase diagram of 5d Yang-Mills theory}},  {\em PoS} {\bf LAT2009} (2009) 050,
  [\href{http://arxiv.org/abs/0911.3609}{{\tt arXiv:0911.3609}}].

\bibitem{Datta:1998eb}
S.~Datta and S.~Gupta, {\it {Dimensional reduction and screening masses in pure
  gauge theories at finite temperature}},  {\em Nucl. Phys.} {\bf B534} (1998)
  392--416, [\href{http://arxiv.org/abs/hep-lat/9806034}{{\tt
  hep-lat/9806034}}].

\bibitem{Chandrasekharan:1996ih}
S.~Chandrasekharan and U.~J. Wiese, {\it {Quantum link models: A discrete
  approach to gauge theories}},  {\em Nucl. Phys.} {\bf B492} (1997) 455--474,
  [\href{http://arxiv.org/abs/hep-lat/9609042}{{\tt hep-lat/9609042}}].

\bibitem{Brower:1997ha}
R.~Brower, S.~Chandrasekharan, and U.~Wiese, {\it {QCD as a quantum link
  model}},  {\em Phys.Rev.} {\bf D60} (1999) 094502,
  [\href{http://arxiv.org/abs/hep-th/9704106}{{\tt hep-th/9704106}}].

\bibitem{Lucini:2004my}
B.~Lucini, M.~Teper, and U.~Wenger, {\it {Glueballs and k-strings in SU(N)
  gauge theories: Calculations with improved operators}},  {\em JHEP} {\bf 06}
  (2004) 012, [\href{http://arxiv.org/abs/hep-lat/0404008}{{\tt
  hep-lat/0404008}}].

\bibitem{Lucini:2010nv}
B.~Lucini, A.~Rago, and E.~Rinaldi, {\it {Glueball masses in the large N
  limit}},  {\em JHEP} {\bf 1008} (2010) 119,
  [\href{http://arxiv.org/abs/1007.3879}{{\tt arXiv:1007.3879}}].

\bibitem{Irges:2006hg}
N.~Irges and F.~Knechtli, {\it {Lattice Gauge Theory Approach to Spontaneous
  Symmetry Breaking from an Extra Dimension}},  {\em Nucl. Phys.} {\bf B775}
  (2007) 283--311, [\href{http://arxiv.org/abs/hep-lat/0609045}{{\tt
  hep-lat/0609045}}].

\bibitem{Luscher:2001up}
M.~Luscher and P.~Weisz, {\it {Locality and exponential error reduction in
  numerical lattice gauge theory}},  {\em JHEP} {\bf 0109} (2001) 010,
  [\href{http://arxiv.org/abs/hep-lat/0108014}{{\tt hep-lat/0108014}}].

\bibitem{Morningstar:1999rf}
C.~J. Morningstar and M.~J. Peardon, {\it {The glueball spectrum from an
  anisotropic lattice study}},  {\em Phys. Rev.} {\bf D60} (1999) 034509,
  [\href{http://arxiv.org/abs/hep-lat/9901004}{{\tt hep-lat/9901004}}].

\end{thebibliography}\endgroup

%
%%%%%%%%% tables %%%%%%%%
%
\TABLE[p]{
  \label{tab:points-n5-4}
  \begin{tabular}[h]{|c|c|c|c|c|c|c|}
  \hline
  \multicolumn{7}{|c|}{$N_4=10$ $N_5=4$}\\
  \hline \hline
  $\beta_4$ & $\beta_5$ & $\beta$ & $\gamma$ & $\frac{N_5}{\gamma}$ &
  $\xi$ & $\frac{N_5}{\xi}$ \\
  \hline
  1.00 & 2.90 & 1.7029 & 1.7029 & 2.3488 & 2.291(39) & 1.744(29) \\
  1.00 & 3.00 & 1.7320 & 1.7320 & 2.3094 & 2.339(40) & 1.709(29) \\
  1.05 & 2.65 & 1.6680 & 1.5886 & 2.5178 & 2.095(39) & 1.911(35) \\
  1.05 & 2.70 & 1.6837 & 1.6035 & 2.4944 & 2.119(38) & 1.889(35) \\
  1.05 & 2.80 & 1.7146 & 1.6329 & 2.4494 & 2.169(36) & 1.844(32) \\
  1.05 & 2.90 & 1.7449 & 1.6619 & 2.4068 & 2.219(37) & 1.801(31) \\
  1.05 & 3.00 & 1.7748 & 1.6903 & 2.3664 & 2.268(38) & 1.764(30) \\
  1.07 & 2.65 & 1.6838 & 1.5737 & 2.5417 & 2.067(43) & 1.934(37) \\
  1.07 & 2.70 & 1.6997 & 1.5885 & 2.5180 & 2.092(40) & 1.911(35) \\
  1.07 & 2.75 & 1.7153 & 1.6031 & 2.4950 & 2.117(40) & 1.889(34) \\
  1.07 & 2.80 & 1.7309 & 1.6176 & 2.4727 & 2.141(38) & 1.866(33) \\
  1.07 & 2.90 & 1.7615 & 1.6462 & 2.4297 & 2.195(38) & 1.824(30) \\
  1.09 & 2.65 & 1.6995 & 1.5592 & 2.5653 & 2.040(43) & 1.959(39) \\
  1.09 & 2.70 & 1.7155 & 1.5738 & 2.5415 & 2.068(41) & 1.934(38) \\
  1.10 & 2.65 & 1.7073 & 1.5521 & 2.5771 & 2.029(42) & 1.969(41) \\
  1.10 & 2.70 & 1.7233 & 1.5667 & 2.5531 & 2.056(41) & 1.948(39) \\
  1.10 & 2.80 & 1.7549 & 1.5954 & 2.5071 & 2.105(38) & 1.900(36) \\
  1.10 & 2.90 & 1.7860 & 1.6236 & 2.4635 & 2.152(37) & 1.858(33) \\
  1.10 & 3.00 & 1.8165 & 1.6514 & 2.4221 & 2.202(38) & 1.815(32) \\
  1.11 & 2.65 & 1.7150 & 1.5451 & 2.5888 & 2.018(44) & 1.986(43) \\
  1.11 & 2.70 & 1.7311 & 1.5596 & 2.5647 & 2.042(42) & 1.959(41) \\
  1.12 & 2.65 & 1.7227 & 1.5382 & 2.6004 & 2.009(44) & 1.998(44) \\
  1.12 & 2.70 & 1.7389 & 1.5526 & 2.5762 & 2.032(42) & 1.972(43) \\
  1.13 & 2.65 & 1.7304 & 1.5313 & 2.6120 & 1.995(45) & 2.008(47) \\
  1.13 & 2.70 & 1.7467 & 1.5457 & 2.5877 & 2.019(44) & 1.984(45) \\
  1.14 & 2.65 & 1.7381 & 1.5246 & 2.6235 & 1.981(45) & 2.019(45) \\
  1.14 & 2.70 & 1.7544 & 1.5389 & 2.5991 & 2.007(45) & 1.995(44) \\
  1.15 & 2.60 & 1.7291 & 1.5036 & 2.6602 & 1.944(47) & 2.059(58) \\
  1.15 & 2.65 & 1.7457 & 1.5180 & 2.6350 & 1.969(45) & 2.031(48) \\
  1.15 & 2.70 & 1.7621 & 1.5322 & 2.6105 & 1.996(43) & 2.006(45) \\
  1.15 & 2.80 & 1.7944 & 1.5603 & 2.5634 & 2.043(40) & 1.955(40) \\
  1.15 & 2.90 & 1.8262 & 1.588  & 2.5188 & 2.092(37) & 1.912(36) \\
  1.15 & 3.00 & 1.8574 & 1.6151 & 2.4765 & 2.139(38) & 1.869(34) \\
  1.20 & 2.50 & 1.7320 & 1.4433 & 2.7712 & 1.839(50) & 2.180(58) \\
  1.20 & 2.55 & 1.7492 & 1.4577 & 2.7439 & 1.861(49) & 2.151(58) \\
  1.20 & 2.60 & 1.7663 & 1.4719 & 2.7174 & 1.886(50) & 2.121(54) \\
  1.20 & 2.65 & 1.7832 & 1.4860 & 2.6917 & 1.912(48) & 2.091(51) \\
  1.20 & 2.70 & 1.8000 & 1.5000 & 2.6666 & 1.939(49) & 2.063(53) \\
  1.10153 & 2.62363 & 1.70 & 1.54 & 2.59 & 2.014(44) & 1.986(42) \\
  1.10801 & 2.63906 & 1.71 & 1.54 & 2.59 & 2.014(44) & 1.986(42) \\
  1.11449 & 2.6545  & 1.72 & 1.54 & 2.59 & 2.014(44) & 1.986(42) \\
  1.12097 & 2.66993 & 1.73 & 1.54 & 2.59 & 2.014(44) & 1.986(42) \\
  1.12745 & 2.68536 & 1.74 & 1.54 & 2.59 & 2.014(44) & 1.986(42) \\
  1.13393 & 2.70079 & 1.75 & 1.54 & 2.59 & 2.014(44) & 1.986(42) \\
  1.14041 & 2.71623 & 1.76 & 1.54 & 2.59 & 2.014(44) & 1.986(42) \\
  1.14688 & 2.73166 & 1.77 & 1.54 & 2.59 & 2.014(44) & 1.986(42) \\
  \hline
\end{tabular}
\caption{This table give the details of the simulated points on the
  $10^4 \times 4$ lattice. Together with the bare parameters
  $\beta_4$, $\beta_5$, $\beta$, $\gamma$ and $\frac{N_5}{\gamma}$, we
  report the values of the renormalised anisotropy $\xi$ from the
  interpolation shown in Fig.~\ref{fig:plottedxi}, and the corresponding
  scale separation $\frac{\LambdaUV}{\LambdaR} = \frac{N_5}{\xi}$.}
}
\TABLE[p]{
  \label{tab:points-n5-6}
  \begin{tabular}[h]{|c|c|c|c|c|c|c|}
  \hline
  \multicolumn{7}{|c|}{$N_4=12$ $N_5=6$}\\
  \hline \hline
  $\beta_4$ & $\beta_5$ & $\beta$ & $\gamma$ & $\frac{N_5}{\gamma}$ &
  $\xi$ & $\frac{N_5}{\xi}$ \\
  \hline
  0.845 & 3.80 & 1.7919 & 2.1206 & 2.8293 & 2.976(30) & 2.016(19) \\
  0.850 & 3.75 & 1.7853 & 2.1004 & 2.8565 & 2.941(30) & 2.039(20) \\
  0.850 & 3.85 & 1.8090 & 2.1282 & 2.8192 & 2.988(31) & 2.008(20) \\
  0.855 & 3.70 & 1.7786 & 2.0802 & 2.8842 & 2.909(27) & 2.061(20) \\
  0.855 & 3.75 & 1.7906 & 2.0942 & 2.8649 & 2.934(29) & 2.045(19) \\
  0.855 & 3.80 & 1.8025 & 2.1081 & 2.8460 & 2.957(29) & 2.029(20) \\
  0.855 & 3.85 & 1.8143 & 2.1220 & 2.8275 & 2.978(31) & 2.016(20) \\
  0.860 & 3.70 & 1.7838 & 2.0742 & 2.8926 & 2.902(29) & 2.069(20) \\
  0.860 & 3.75 & 1.7958 & 2.0881 & 2.8733 & 2.923(29) & 2.053(20) \\
  0.860 & 3.80 & 1.8077 & 2.1020 & 2.8543 & 2.945(28) & 2.038(20) \\
  0.860 & 3.85 & 1.8196 & 2.1158 & 2.8357 & 2.968(29) & 2.022(20) \\
  0.865 & 3.55 & 1.7523 & 2.0258 & 2.9617 & 2.823(32) & 2.126(23) \\
  0.865 & 3.60 & 1.7646 & 2.0400 & 2.9410 & 2.846(31) & 2.109(23) \\
  0.865 & 3.65 & 1.7768 & 2.0541 & 2.9208 & 2.869(29) & 2.092(21) \\
  0.865 & 3.70 & 1.7889 & 2.0682 & 2.9010 & 2.889(29) & 2.076(21) \\
  0.865 & 3.80 & 1.8130 & 2.0959 & 2.8626 & 2.936(30) & 2.045(21) \\
  0.870 & 3.55 & 1.7574 & 2.0200 & 2.9702 & 2.811(32) & 2.133(24) \\
  0.870 & 3.60 & 1.7697 & 2.0341 & 2.9495 & 2.836(32) & 2.117(23) \\
  0.870 & 3.65 & 1.7819 & 2.0482 & 2.9293 & 2.858(30) & 2.099(23) \\
  0.870 & 3.70 & 1.7941 & 2.0622 & 2.9094 & 2.880(29) & 2.083(22) \\
  0.875 & 3.55 & 1.7624 & 2.0142 & 2.9788 & 2.802(33) & 2.140(24) \\
  0.875 & 3.60 & 1.7748 & 2.0283 & 2.9580 & 2.823(31) & 2.123(24) \\
  0.875 & 3.65 & 1.7871 & 2.0424 & 2.9377 & 2.849(31) & 2.106(22) \\
  0.875 & 3.70 & 1.7993 & 2.0563 & 2.9177 & 2.872(29) & 2.090(21) \\
  0.880 & 3.55 & 1.7674 & 2.0085 & 2.9873 & 2.792(34) & 2.149(25) \\
  0.888 & 3.50 & 1.7629 & 1.9853 & 3.0222 & 2.755(34) & 2.179(26) \\
  0.890 & 3.55 & 1.7775 & 1.9971 & 3.0042 & 2.775(35) & 2.163(27) \\
  0.900 & 3.45 & 1.7621 & 1.9578 & 3.0645 & 2.711(31) & 2.212(26) \\
  0.900 & 3.50 & 1.7748 & 1.9720 & 3.0425 & 2.735(32) & 2.194(27) \\
  0.900 & 3.55 & 1.7874 & 1.9860 & 3.0210 & 2.756(33) & 2.178(26) \\
  0.908 & 3.45 & 1.7699 & 1.9492 & 3.0781 & 2.699(33) & 2.226(25) \\
  0.920 & 3.40 & 1.7686 & 1.9224 & 3.1210 & 2.653(31) & 2.261(26) \\
  0.920 & 3.45 & 1.7815 & 1.9364 & 3.0983 & 2.676(32) & 2.241(26) \\
  0.920 & 3.50 & 1.7944 & 1.9504 & 3.0761 & 2.699(30) & 2.223(26) \\
  0.860 & 3.60 & 1.7595 & 2.0459 & 2.9325 & 2.855(30) & 2.102(22) \\
  0.880 & 3.60 & 1.7798 & 2.0226 & 2.9664 & 2.816(33) & 2.131(25) \\
  0.900 & 3.60 & 1.8000 & 2.0000 & 3.0000 & 2.777(34) & 2.159(26) \\
  0.920 & 3.55 & 1.8072 & 1.9643 & 3.0544 & 2.722(33) & 2.205(26) \\
  0.920 & 3.60 & 1.8198 & 1.9781 & 3.0331 & 2.743(32) & 2.187(27) \\
  0.940 & 3.40 & 1.7877 & 1.9018 & 3.1548 & 2.619(30) & 2.291(26) \\
  0.940 & 3.50 & 1.8138 & 1.9296 & 3.1094 & 2.665(30) & 2.251(26) \\
  0.940 & 3.60 & 1.8395 & 1.9569 & 3.0659 & 2.709(31) & 2.215(25) \\
  \hline
\end{tabular}
\caption{This table give the details of the simulated points on the
  $12^4 \times 6$ lattice. Together with the bare parameters
  $\beta_4$, $\beta_5$, $\beta$, $\gamma$ and $\frac{N_5}{\gamma}$, we
  report the values of the renormalised anisotropy $\xi$ from the
  interpolation shown in Fig.~\ref{fig:plottedxi}, and the corresponding
  scale separation $\frac{\LambdaUV}{\LambdaR} = \frac{N_5}{\xi}$.}
}
\newpage
\clearpage
\begin{table}[p]
  \centering
  \begin{tabular}[c]{|c|c|c|c|}
    \hline
    \multicolumn{4}{|c|}{Scalar masses on $N_4=10$ $N_5=4$} \\
    \hline \hline
    $\beta_4$ & $\beta_5$ & $a_4m_5$ & $t_{\rm min}-t_{\rm max}$ \\
    \hline
    1.00 & 2.90 & 1.118(26) & 2 - 4 \\
    1.00 & 3.00 & 1.406(56) & 2 - 4 \\
    1.00 & 3.00 & 1.30(19)  & {\bf 3 - 4} \\
    1.05 & 2.80 & 1.020(16) & 2 - 4 \\
    1.05 & 3.00 & 1.386(85) & 2 - 4 \\
    1.05 & 3.00 & 1.05(22)  & {\bf 3 - 4} \\ 
    1.07 & 2.70 & 0.6482(69)& 2 - 4 \\
    1.07 & 2.75 & 0.925(13) & 2 - 4 \\
    1.07 & 2.75 & 0.904(29) & {\bf 3 - 4} \\
    1.07 & 2.80 & 1.022(57) & 3 - 4 \\
    1.07 & 2.90 & 1.349(56) & 2 - 4 \\
    1.07 & 2.90 & 1.27(21)  & {\bf 3 - 4} \\
    1.09 & 2.65 & 0.4586(43)& 2 - 4 \\
    1.09 & 2.70 & 0.844(10) & 2 - 4 \\
    1.10 & 2.65 & 0.624(05) & 2 - 4 \\
    1.10 & 2.70 & 0.880(14) & 2 - 4 \\
    1.10 & 2.80 & 1.258(51) & 2 - 4 \\
    1.10 & 2.80 & 1.15(15) & {\bf 3 - 4} \\
    1.10 & 2.90 & 1.64(17) & 2 - 4 \\
    1.11 & 2.65 & 0.732(11) & 2 - 4 \\
    1.11 & 2.70 & 0.865(32) & 3 - 4 \\
    1.12 & 2.65 & 0.779(11) & 2 - 4 \\
    1.14 & 2.65 & 0.840(52) & 3 - 4 \\
    1.14 & 2.70 & 0.90(13)  & 3 - 4 \\
    1.15 & 2.60 & 0.724(21) & 3 - 4 \\
    1.15 & 2.70 & 1.424(55) & 2 - 4 \\
    1.15 & 2.70 & 1.24(21) & {\bf 3 - 4} \\
    1.20 & 2.50 & 0.923(76) & 3 - 4 \\
    1.10801 & 2.63906 & 0.638(44) & 3 - 4 \\
    1.11449 & 2.6545  & 0.765(12) & 3 - 4 \\
    1.12097 & 2.66993 & 0.822(26) & 3 - 4 \\
    1.12745 & 2.68536 & 0.898(37) & 3 - 4 \\
    1.13393 & 2.70079 & 0.791(98) & 3 - 4 \\
    1.14041 & 2.71623 & 1.450(92) & 2 - 4 \\
    1.14041 & 2.71623 & 1.18(34)  & {\bf 3 - 4} \\
    \hline
  \end{tabular}
  \label{tab:scalars-n5-4}
  \caption{Measured scalar masses on the $10^4 \times 4$ lattice. The
    fitting range for the effective mass plateaux is shown in the last
    column. Boldface values are alternative fitting ranges.}
\end{table}
\clearpage
\newpage
\TABLE[p]{
  \label{tab:torelons-n5-4}
  \begin{tabular}[h]{|c|c|c|c|c|c|}
    \hline
    \multicolumn{6}{|c|}{Torelon masses and string tensions on $N_4=10$ $N_5=4$} \\
    \hline \hline
    $\beta_4$ & $\beta_5$ & $a_4m_{\rm tor}$ & $t_{\rm min}-t_{\rm max}$ & $a_4\sqrt{\sigma}$ & $\sim L_4\sqrt{\sigma}$\\
    \hline
    1.05 & 3.00 & 1.219(24)  & 2 - 4  &  0.3638(33) & 3.6 \\
    1.05 & 3.00 & 1.173(71)  & {\bf 3 - 4} & 0.3574(99) & 3.6 \\
    1.07 & 2.90 & 1.334(39)  & 2 - 4  &  0.3793(52) & 3.8 \\
    1.07 & 2.90 & 1.16(11)  & {\bf 3 - 4}  & 0.356(15) & 3.6 \\
    1.10 & 2.80 & 1.116(16)  & 2 - 4  &  0.3493(23) & 3.5 \\
    1.10 & 2.80 & 1.105(48)  & {\bf 3 - 4}  & 0.3478(69) & 3.5 \\
    1.10 & 2.90 & 0.428(2)  & 2 - 4  &  0.2308(5) & 2.3 \\
    1.14 & 2.65 & 1.215(27)  & 2 - 4  &  0.3633(37) & 3.6 \\
    1.14 & 2.65 & 1.190(71)  & {\bf 3 - 4} & 0.3598(99) & 3.6 \\
    1.14 & 2.70 & 0.6153(39) & 2 - 4  &  0.2683(7) & 2.7 \\
    1.15 & 2.70 & 0.3798(14) & 2 - 4  &  0.2201(3) & 2.2 \\
    1.20 & 2.50 & 0.5663(41) & 2 - 4  &  0.2590(8) & 2.6 \\
    1.12745 & 2.68536 & 1.18(10)   &  3 - 4 &  0.358(14)  & 3.6 \\
    1.13393 & 2.70079 & 0.8175(78) &  2 - 4 &  0.3037(13) & 3.0 \\
    1.14041 & 2.71623 & 0.491(3) &  2 - 3 &  0.2441(6)  & 2.4 \\
    \hline
  \end{tabular}
  \caption{Measured torelon masses on the $10^4 \times 4$ lattice. The
    fitting range for the effective mass plateaux is shown together with
    the calculated string tension. Boldface values are alternative fitting ranges.
    The last column states the spatial length of the lattice in units of the
    four dimensional correlation length given by the inverse of the
    string tension.}
}
\newpage
\clearpage
\begin{table}[p]
  \centering
  \begin{tabular}[h]{|c|c|c|c|}
    \hline
    \multicolumn{4}{|c|}{Scalar masses on $N_4=12$ $N_5=6$}\\
    \hline \hline
    $\beta_4$ & $\beta_5$ & $a_4m_5$ & $t_{\rm min}-t_{\rm max}$ \\
    \hline
    0.845 & 3.80 & 1.05(31)   & 3 - 5 \\
    0.845 & 3.80 & {\bf 1.17(20)} & 3 - 5 \\
    0.850 & 3.75 & 1.273(74)  & 2 - 3 \\
    0.850 & 3.75 & 1.14(19)   & {\bf 3 - 4} \\
    0.850 & 3.85 & 1.65(15)   & 2 - 4 \\
    0.850 & 3.85 & 1.04(36)   & {\bf 3 - 4} \\
    0.855 & 3.70 & 0.91(14)   & 3 - 5 \\
    0.855 & 3.75 & 1.35(25)   & 3 - 5 \\
    *0.860 & 3.60 & 0.581(20)  & 4 - 7 \\
    0.860 & 3.70 & 1.294(67)  & 2 - 4 \\
    0.860 & 3.70 & 1.19(24) & {\bf 3 - 4} \\
    0.865 & 3.55 & 0.3312(38) & 2 - 4 \\
    0.865 & 3.60 & 0.677(25)  & 4 - 5 \\
    0.865 & 3.65 & 0.95(10)   & 3 - 5 \\
    0.870 & 3.55 & 0.4679(86) & 4 - 5 \\
    0.870 & 3.60 & 0.818(58)  & 3 - 4 \\
    0.870 & 3.65 & 0.998(93)  & 3 - 5 \\
    0.870 & 3.70 & 1.401(92)   & 2 - 4 \\
    0.870 & 3.70 & 1.07(26)   & {\bf 3 - 4} \\ 
    0.875 & 3.55 & 0.532(20)  & 4 - 5 \\
    0.875 & 3.60 & 0.880(97)  & 3 - 4 \\
    0.875 & 3.65 & 1.40(8)   & 2 - 4 \\
    0.875 & 3.65 & 1.16(25)   & {\bf 3 - 4} \\
    0.875 & 3.70 & 1.02(20)   & 3 - 5 \\
    0.880 & 3.55 & 0.727(50)  & 3 - 5 \\
    0.880 & 3.55 & {\bf 0.86(15)}  & 3 - 5 \\
    *0.880 & 3.60 & 1.318(57)  & 2 - 4 \\
    *0.880 & 3.60 & 1.30(21)   & {\bf 3 - 4} \\
    0.888 & 3.50 & 0.595(19)  & 3 - 5 \\
    0.888 & 3.50 & 0.561(34)  & {\bf 4 - 5} \\
    0.890 & 3.55 & 1.038(82)  & 3 - 5 \\
    0.890 & 3.55 & 0.93(21)   & {\bf 4 - 5} \\
    0.900 & 3.45 & 0.635(12)  & 2 - 5 \\
    0.900 & 3.50 & 1.173(72)  & 2 - 5 \\
    0.900 & 3.50 & 0.91(13)   & {\bf 4 - 5} \\
    0.908 & 3.45 & 0.777(78)  & 3 - 5 \\
    0.920 & 3.45 & 1.46(11)   & 2 - 4 \\
    0.920 & 3.45 & 1.06(33)   & {\bf 3 - 4} \\
    \hline
  \end{tabular}
  \label{tab:scalars-n5-6}
  \caption{Measured scalar masses on the $12^4 \times 6$ lattice. The
    fitting range for the effective mass plateaux is shown in the last
    column. Boldface values are alternative fitting ranges. Boldface
    masses come from a different choice of the scalar operator in the
    correlator. (The starred points
    come from a lattice with a longer temporal distance $L_t=2L_4$)}
\end{table}
\clearpage
\newpage
\TABLE[p]{
  \label{tab:torelons-n5-6}
  \begin{tabular}[h]{|c|c|c|c|c|c|}
    \hline
    \multicolumn{6}{|c|}{Torelon masses and string tensions on $N_4=12$ $N_5=6$} \\
    \hline \hline
    $\beta_4$ & $\beta_5$ & $a_4m_{\rm tor}$ & $t_{\rm min}-t_{\rm max}$ & $a_4\sqrt{\sigma}$ & $\sim L_4\sqrt{\sigma}$\\
    \hline
    0.845 & 3.80 & 0.5800(60) & 2 - 5 & 0.2358(11) & 2.8 \\
    0.850 & 3.75 & 0.7127(75) & 2 - 5 & 0.2582(12) & 3.1 \\
    0.850 & 3.75 & 0.730(16) & {\bf 3 - 5} & 0.2610(26) & 3.1 \\
    0.850 & 3.85 & 0.2665(19) & 2 - 4 & 0.1717(4) & 2.1 \\
    0.850 & 3.85 & 0.2658(18) & {\bf 3 - 4} & 0.1715(4) & 2.1 \\
    0.855 & 3.70 & 0.8954(99) & 2 - 5 &  0.2862(14) & 3.4 \\
    0.855 & 3.75 & 0.5162(51) & 2 - 5 & 0.2243(9) & 2.7 \\
    0.860 & 3.70 & 0.6412(73) & 2 - 5 & 0.2464(12) & 3.0 \\
    0.865 & 3.65 & 0.808(19)  & 3 - 5 & 0.2732(29) & 3.3 \\
    0.865 & 3.65 & 0.784(34)  & {\bf 4 - 5} & 0.2694(53) & 3.2 \\
    0.870 & 3.60 & 1.224(77)  & 3 - 5 & 0.3306(97) & 4.0 \\
    0.870 & 3.65 & 0.5798(62) & 2 - 5 & 0.2358(11) & 2.8 \\
    0.870 & 3.70 & 0.3295(19) & 2 - 5 & 0.1864(4) & 2.2 \\
    0.875 & 3.60 & 0.7933(92) & 2 - 4 & 0.2709(14) & 3.3 \\
    0.875 & 3.65 & 0.4177(31) & 2 - 5 & 0.2051(6) & 2.5 \\
    0.875 & 3.70 & 0.2418(10) & 3 - 5 & 0.1656(2) & 2.0 \\
    0.880 & 3.55 & 1.25(12)   & 3 - 5 & 0.334(14) & 4.0 \\
    *0.880 & 3.60 & 0.523(21)  & 5 - 11 & 0.2256(40) & 2.7 \\
    0.890 & 3.55 & 0.5140(49) & 2 - 5 & 0.2239(9) & 2.7 \\
    0.900 & 3.45 & 1.06(11)   & 3 - 5 & 0.310(15) & 3.7 \\
    0.900 & 3.45 & 0.87(25)   & {\bf 4 - 5} & 0.282(37) & 3.4 \\
    0.900 & 3.50 & 0.4963(52) & 2 - 5 & 0.2205(10) & 2.6 \\
    0.908 & 3.45 & 0.5942(67) & 2 - 5 & 0.2383(12) & 2.9 \\
    0.920 & 3.45 & 0.2353(19) & 2 - 4 & 0.1639(5) & 2.0 \\
    \hline
  \end{tabular}
  \caption{Measured torelon masses on the $12^4 \times 6$ lattice. The
    fitting range for the effective mass plateaux is shown together with
    the calculated string tension. Boldface values are alternative fitting ranges.
    The last column states the spatial length of the lattice in units of the
    four dimensional correlation length given by the inverse of the string tension.
    (The starred point comes from a lattice with a longer temporal distance $L_t=2L_4$)}
}
\TABLE[p]{
  \begin{tabular}[h]{|c|c|c|c|c|}
    \hline
    \multicolumn{5}{|c|}{$N_4=10$ $N_5=4$} \\
    \hline \hline
    $\frac{\LambdaUV}{\LambdaR}$ & $a_4\sqrt{\sigma}$ & $R\sqrt{\sigma}$ & $\frac{m_5}{\sqrt{\sigma}}$ & $Rm_5$ \\
    \hline
    1.763(29) & 0.3638(33) & 0.1021(19) & 3.81(23) & 0.389(25) \\% 3.63772 1.05 3.00 
    1.763(29) & 0.3638(33) & 0.1021(19) & 2.89(61) & 0.295(63) \\% 3.63772
    1.763(29) & 0.3574(99) & 0.1003(32) & 3.88(26) & 0.389(29) \\% 3.57419
    1.763(29) & 0.3574(99) & 0.1003(32) & 2.95(62) & 0.295(63) \\% 3.57419
    \hline
    1.824(30) & 0.3793(52) & 0.1101(23) & 3.56(15) & 0.392(19) \\% 3.79261 1.07 2.90 
    1.824(30) & 0.3793(52) & 0.1101(23) & 3.35(54) & 0.369(60) \\% 3.79261
    1.824(30) & 0.356(15) & 0.1033(48) & 3.79(21) & 0.392(28)  \\% 3.55704
    1.824(30) & 0.356(15) & 0.1033(48) & 3.57(57) & 0.369(62)  \\% 3.55704
    \hline
    1.900(35) & 0.3493(23) & 0.1056(21) & 3.60(14) & 0.380(17) \\% 3.49322 1.10 2.80 
    1.900(35) & 0.3493(23) & 0.1056(21) & 3.29(41) & 0.348(44) \\% 3.49322
    1.900(35) & 0.3478(69) & 0.1052(29) & 3.62(16) & 0.380(20) \\% 3.47793
    1.900(35) & 0.3478(69) & 0.1052(29) & 3.31(41) & 0.348(44) \\% 3.47793
    \hline
    1.857(32) & 0.2308(5) & 0.0682(12) & 7.10(75) & 0.484(52) \\% 2.30796 1.10 2.90 
    \hline
    2.019(45) & 0.3633(37) & 0.1168(29) & 2.31(15) & 0.270(18) \\% 3.63329 1.14 2.65 
    2.019(45) & 0.3598(99) & 0.1156(41) & 2.33(16) & 0.270(20) \\% 3.59768
    \hline
    1.994(43) & 0.2683(7) & 0.0852(19) & 3.36(50) & 0.286(43) \\% 2.68333 1.14 2.70 
    \hline
    2.006(44) & 0.2201(33) & 0.0703(15) & 6.47(24) & 0.455(20) \\% 2.20126 1.15 2.70 
    2.006(44) & 0.2201(33) & 0.0703(15) & 5.66(94) & 0.397(67) \\% 2.20126
    \hline
    2.180(57) & 0.2590(8) & 0.0899(24) & 3.56(30) & 0.320(28) \\% 2.59048 1.20 2.50 
    \hline
    1.986(42) & 0.358(14) & 0.1133(51) & 2.51(14) & 0.284(20) \\%3.5848 1.12745 2.68536 
    \hline
    1.986(42) & 0.3037(13) & 0.0960(21) & 2.61(32) & 0.250(31) \\%3.03675 1.13393 2.70079 
    \hline
    1.986(42) & 0.2441(6) & 0.0772(16) & 5.94(38) & 0.458(31) \\%2.4412 1.14041 2.71623 
    1.986(42) & 0.2441(6) & 0.0772(16) & 4.9(1.4) & 0.37(11) \\%2.4412
    \hline
  \end{tabular}
  \label{tab:results-n5-4}
  \caption{The results of our numerical simulations are reported in
    this table for $N_5=4$. The primary observables directly measured in the
    simulations, 
    $a_4\sqrt{\sigma}$ and $a_4m_5$, are used to obtain the
    combinations $\frac{m_5}{\sqrt{\sigma}}$, $R\sqrt{\sigma}$ and
    $Rm_5$. This is done using the value for the separation of the
    cut--off scale from the compactification scale
    $\frac{\LambdaUV}{\LambdaR}$. When more than one value is shown,
    they come from different fitting ranges in the plateaux of the
    primary observables. The spread of the values is used to estimate
    a systematic error that is then reported in the plots.}
}

\TABLE[p]{
  \begin{tabular}[h]{|c|c|c|c|c|}
    \hline
    \multicolumn{5}{|c|}{$N_4=12$ $N_5=6$} \\
    \hline \hline
    $\frac{\LambdaUV}{\LambdaR}$ & $a_4\sqrt{\sigma}$ & $R\sqrt{\sigma}$ & $\frac{m_5}{\sqrt{\sigma}}$ & $Rm_5$ \\
    \hline
    2.016(19) & 0.2358(11) & 0.0757(8) & 4.5(1.3) & 0.337(99) \\% 2.82955 0.845 3.80 
    2.016(19) & 0.2358(11) & 0.0757(8) & 4.98(84) & 0.377(64) \\% 2.82955
    \hline
    2.039(20) & 0.2582(12) & 0.0838(9) & 4.93(29) & 0.413(25) \\% 3.09833 0.85 3.75 
    2.039(20) & 0.2610(26) & 0.0847(12) & 4.88(29) & 0.413(25) \\% 3.13248
    2.039(20) & 0.2582(12) & 0.0838(9) & 4.41(75) & 0.369(63) \\% 3.09833
    2.039(20) & 0.2610(26) & 0.0847(12) & 4.36(74) & 0.369(63) \\% 3.13248
    \hline
    2.008(20) & 0.1717(5) & 0.0549(6) & 9.59(88) & 0.526(49) \\% 2.06032 0.85 3.85 
    2.008(20) & 0.1715(4) & 0.0548(6) & 9.60(88) & 0.526(49) \\% 2.05846
    2.008(20) & 0.1717(5) & 0.0549(6) & 6.1(2.1) & 0.33(12) \\% 2.06032
    2.008(20) & 0.1715(4) & 0.0548(6) & 6.1(2.1) & 0.33(12) \\% 2.05846
    \hline
    2.061(19) & 0.2862(14) & 0.0939(10) & 3.20(48) & 0.300(45) \\% 3.434 0.855 3.70 
    \hline
    2.045(19) & 0.2243(9) & 0.0730(8) & 6.0(1.1) & 0.439(80) \\% 2.69112 0.855 3.75 
    \hline
    2.069(20) & 0.2464(12) & 0.0811(9) & 5.25(28) & 0.426(23) \\% 2.95667 0.86 3.70 
    2.069(20) & 0.2464(12) & 0.0811(9) & 4.84(98) & 0.393(79) \\% 2.95667
    \hline
    2.092(21) & 0.2732(29) & 0.0910(13) & 3.49(38) & 0.317(35) \\% 3.2778 0.865 3.65 
    2.092(21) & 0.2694(53) & 0.0897(20) & 3.54(39) & 0.317(35) \\% 3.23278
    \hline
    2.117(23) & 0.3306(97) & 0.1114(35) & 2.47(19) & 0.276(23) \\% 3.96722 0.870 3.60 
    \hline
    2.099(23) & 0.2358(11) & 0.0788(9) & 4.23(39) & 0.333(31) \\% 2.82935 0.870 3.65 
    \hline
    2.083(22) & 0.1864(41) & 0.0618(7) & 7.52(50) & 0.464(31) \\% 2.23624 0.870 3.70 
    2.083(22) & 0.1864(41) & 0.0618(7) & 5.7(1.4) & 0.355(86) \\% 2.23624
    \hline
    2.123(24) & 0.2709(14) & 0.0915(11) & 3.25(36) & 0.297(33) \\% 3.25073 0.875 3.60 
    \hline
    2.106(22) & 0.2051(64) & 0.0688(7) & 6.83(41) & 0.469(29) \\% 2.46163 0.875 3.65 
    2.106(22) & 0.2051(64) & 0.0688(7) & 5.6(1.2) & 0.388(84) \\% 2.46163
    \hline
    2.090(21) & 0.1656(2) & 0.0551(5) & 6.2(1.2) & 0.339(65) \\% 1.98713  0.875 3.70
    \hline
    2.149(25) & 0.334(14) & 0.1144(51) & 2.18(18) & 0.249(24) \\% 4.01288 0.88 3.55 
    2.149(25) & 0.334(14) & 0.1144(51) & 2.57(45) & 0.294(53) \\% 4.01288
    \hline
    2.131(25) & 0.2256(40) & 0.0765(16) & 5.84(25) & 0.447(21) \\% 2.70715 0.88 3.60 
    2.131(25) & 0.2256(40) & 0.0765(16) & 5.76(92) & 0.441(71) \\% 2.70715
    \hline
    2.163(27) & 0.2239(9) & 0.0771(10) & 4.64(37) & 0.357(29) \\% 2.68621 0.89 3.55 
    2.163(27) & 0.2239(9) & 0.0771(10) & 4.16(93) & 0.320(72) \\% 2.68621
    \hline
    2.212(26) & 0.310(15) & 0.1090(54) & 2.05(11) & 0.224(16) \\% 3.71593 0.90 3.45 
    2.212(26) & 0.282(37) & 0.099(13) & 2.25(31) & 0.224(42) \\% 3.38533
    \hline
    2.194(27) & 0.2205(10) & 0.0770(10) & 5.32(33) & 0.410(26) \\% 2.64632 0.90 3.50 
    2.194(27) & 0.2205(10) & 0.0770(10) & 4.12(60) & 0.317(46) \\% 2.64632
    \hline
    2.226(25) & 0.2383(12) & 0.0844(10) & 3.26(33) & 0.275(28) \\% 2.85965 0.908 3.45 
    \hline
    2.241(26) & 0.1639(5) & 0.0585(69) & 8.90(66) & 0.520(39) \\% 1.96728 0.92 3.45 
    2.241(26) & 0.1639(5) & 0.0585(69) & 6.4(2.0) & 0.38(12) \\% 1.96728
    \hline
  \end{tabular}
  \label{tab:results-n5-6}
  \caption{The results of our numerical simulations are reported in
    this table for $N_5=6$. The primary observables directly measured in the
    simulations, 
    $a_4\sqrt{\sigma}$ and $a_4m_5$, are used to obtain the
    combinations $\frac{m_5}{\sqrt{\sigma}}$, $R\sqrt{\sigma}$ and
    $Rm_5$. This is done using the value for the separation of the
    cut--off scale from the compactification scale
    $\frac{\LambdaUV}{\LambdaR}$. When more than one value is shown,
    they come from different fitting ranges in the plateaux of the
    primary observables. The spread of the values is used to estimate
    a systematic error that is then reported in the plots.}
}

\end{document}